\documentclass[twocolumn]{aastex631}

\submitjournal{ApJ}

\shortauthors{Williams et al.}

\usepackage[caption=false]{subfig}
\usepackage{mwe}
\usepackage{multirow}
\usepackage{amsmath}
\usepackage{graphicx}

\begin{document}

\title{Observing compact Pop III star clusters and the presence of cosmic streaming}

\correspondingauthor{Claire E. Williams}

\author[0000-0003-2369-2911]{Claire E. Williams}
\affil{Department of Physics and Astronomy, UCLA, Los Angeles, CA 90095}
\affil{Mani L. Bhaumik Institute for Theoretical Physics, Department of Physics and Astronomy, UCLA, Los Angeles, CA 90095, USA\\}
\email{clairewilliams@astro.ucla.edu}

\author[0000-0002-9802-9279]{Smadar Naoz}
\affil{Department of Physics and Astronomy, UCLA, Los Angeles, CA 90095}
\affil{Mani L. Bhaumik Institute for Theoretical Physics, Department of Physics and Astronomy, UCLA, Los Angeles, CA 90095, USA\\}

\author[0000-0001-7925-238X]{Naoki Yoshida}
\affiliation{Department of Physics, The University of Tokyo, 7-3-1 Hongo, Bunkyo, Tokyo 113-0033, Japan}
\affiliation{Kavli Institute for the Physics and Mathematics of the Universe (WPI), UT Institute for Advanced Study, The University of Tokyo, Kashiwa, Chiba 277-8583, Japan}
\affiliation{Research Center for the Early Universe, School of Science, The University of Tokyo, 7-3-1 Hongo, Bunkyo, Tokyo 113-0033, Japan}

\author[0000-0002-4227-7919]{William Lake}
\affil{Department of Physics and Astronomy, Dartmouth College, Hanover, NH 03755, USA \\}

\author[0000-0001-5817-5944]{Blakesley Burkhart}
\affiliation{Department of Physics and Astronomy, Rutgers, The State University of New Jersey, 136 Frelinghuysen Rd, Piscataway, NJ 08854, USA \\}
\affiliation{Center for Computational Astrophysics, Flatiron Institute, 162 Fifth Avenue, New York, NY 10010, USA \\}

\author[0000-0003-3816-7028]{Federico Marinacci}
\affiliation{Department of Physics \& Astronomy ``Augusto Righi", University of Bologna, via Gobetti 93/2, 40129 Bologna, Italy\\}

\author[0000-0001-8593-7692]{Mark Vogelsberger}
\affil{Department of Physics and Kavli Institute for Astrophysics and Space Research, Massachusetts Institute of Technology, Cambridge, MA 02139, USA}
\affil{Fachbereich Physik, Philipps Universit\"at Marburg, D-35032 Marburg, Germany}

\author[0000-0001-5944-291X]{Shyam H. Menon}
\affiliation{Center for Computational Astrophysics, Flatiron Institute, 162 Fifth Avenue, New York, NY 10010, USA \\}
\affiliation{Department of Physics and Astronomy, Rutgers, The State University of New Jersey, 136 Frelinghuysen Rd, Piscataway, NJ 08854, USA \\}

\author[0000-0002-8859-7790]{Avi Chen}
\affiliation{Department of Physics and Astronomy, Rutgers, The State University of New Jersey, 136 Frelinghuysen Rd, Piscataway, NJ 08854, USA \\}

\author[0000-0002-4086-3180]{Kyle Kremer}
\affiliation{Department of Astronomy and Astrophysics, University of California, San Diego, 9500 Gilman Drive, La Jolla, CA 92093, USA}



\begin{abstract}

The formation of 
the Universe's first luminous stellar structures depends on the unique conditions at ``Cosmic Dawn," which are set by the underlying cosmological model and early baryonic physics. 
Observations suggest that high-$z$ star clusters reached stellar surface densities above $10^5 M_\odot$ pc$^{-2}$, suggesting scenarios
where models predict that the ability of stellar feedback to counter gravitational collapse is severely limited.
We investigate the first star clusters in a suite of {\tt AREPO} simulations, which explore the capacity for $\Lambda$CDM halos to maximally form high-density systems without feedback.
We include the effects of the supersonic baryon-dark matter streaming velocity, an effect that impacts gas density and distribution in early minihalos.
We show that early star clusters can reach high densities even in regions of strong supersonic streaming, provided
feedback is weak.
We analyze the interplay of the stream velocity and the dynamical processes of structure formation, 
finding that {\it JWST}
has the opportunity to detect the brightest, most massive objects in our computational box. 
The detection of individual  $z\geq12$ Pop III star clusters below $10^7M_\odot$ is challenging, although
lensing could reveal these objects in rare configurations, especially if a top-heavy IMF is present. 
We find that accounting for baryonic clusters separately from dark matter halos complicates predictions for the faint-end of the high-$z$ UVLF, with competing effects from the stream velocity and low-mass clusters outside of halos. 
Finally, we explore clustering of star clusters  as a promising probe of the stream velocity in these systems. 
\end{abstract}

\keywords{Star clusters --- hydrodynamical simulations --- Population III stars --- James Webb Space Telescope --- high-redshift galaxies}

\section{Introduction} \label{sec:intro}
Understanding the nature of star formation accross the Universe’s 13.8 billion year history is a focus of scientific investigation with the {\it James Webb Space Telescope (JWST)}. 
The specifics of star formation at the earliest times have been historically challenging to untangle due to observational hurdles, leaving room for theoretical uncertainties. 
However, the influx of data from {\it JWST} has revolutionized our ability to probe the first stars and galaxies. 
Now, it is possible not only to observe the emission of galaxies in the first several hundred Myr of the Universe’s existence \citep[e.g.,][]{Finkelstein+23ceersI}, but also to probe individual star clusters through gravitational lensing up to $z\sim 10$ \citep[e.g.,][]{Adamo+24, Mowla+24, pascale_is_2025}.
Some intriguing  structures may even show direct evidence of hosting the Pop III stars \citep[e.g.,][]{fujimoto_glimpse_2025, cullen_ultraviolet_2024, maiolino_jwst-jades_2024}.

These observations paint an unprecedented picture of star formation in the earliest epochs: some high redshift systems are extremely blue, compact, and bright, properties that are challenging to produce with the standard, pre-{\it JWST} galaxy formation paradigm \citep[e.g.,][]{harikane_jwst_2024,adams_discovery_2023,Casey+24, baggen_small_2024,2025arXiv250219484J}.  
Some studies even suggested that modified cosmology may play a role in generating the observed galaxy properties at cosmic dawn \citep[e.g.,][]{Menci_2024_excess}.
Supernova, radiative, stellar wind, and other feedback processes typically prevent the efficient conversion of baryons to stars in collapsing clouds with stellar surface densities around $\Sigma\sim10^{1-3} M_\odot$ pc$^{-2}$ for typical young massive clusters, while globular clusters, nuclear star clusters, and ultra compact dwarf galaxies may reach $\Sigma\sim10^{3-5} M_\odot$ pc$^{-2}$ \citep[e.g.,][]{grudic_maximum_2019, hopkins_maximum_2010}. 
Meanwhile, some high redshift star clusters and galaxies display surface densities of $\Sigma\sim10^{5-6} M_\odot $ pc$^{-2}$ \citep[e.g.,][]{Adamo+24}.
These immense stellar surface densities imply that the processes leading to molecular cloud disruption were impeded, allowing for star formation to proceed unimpeded by feedback. 

The early Universe hosted unique conditions compared to typical star formation environments  today. 
The first ``Population III” (Pop III) stars arose from the primordial gas of the Big Bang, and thus were entirely composed of Hydrogen and Helium, seeding the surrounding medium with the Universe's first metals upon their evolution and collapse. 
The cooling processes leading to collapse in these first low-mass halos must have been atomic and molecular hydrogen cooling alone \citep[e.g.,][]{Yoshida+07}.
This pristine mode of star formation may have persisted in pockets of the universe even as Pop II star formation began elsewhere \citep[see e.g.,][]{hegde_efficient_2025}.
Many studies suggest that the absence of metals allowed Pop III stars to grow to extremely massive size, and that their initial mass function may have been top heavy compared to higher metallicity distributions \citep[e.g.,][]{wollenberg_formation_2020,Chon+22,fukushima_formation_2023,Lake+24b,chon_impact_2024}. 
While massive stars are typically synonymous with strong winds, outflows, and rapid supernovae, theoretical work hints that for systems that achieve surface densities upwards of $10^5M_\odot$ pc$^{-2}$, the gravitational potential of the cloud dominates over disrupting feedback effects, even for top-heavy IMF scenarios \citep[e.g.,][]{grudic_when_2018,lancaster_star_2021,menon_outflows_2023,Menon+24,nebrin_lyman-_2025}. 
This leads to the suggestion by \cite{dekel_efficient_2023} that in many cases, conditions in the gas at these times are ripe for “feedback-free” star formation, where a highly efficient burst of star formation converts gas clouds into stars with virtually no feedback disruption \citep[e.g.,][]{dekel_efficient_2023,Li_2024_Feedback,dekel_growth_2024}.
While \cite{dekel_efficient_2023} suggest that conditions for such feedback-free star formation may occur with stellar mass $M_*\approx 10^{10} M_\odot$ at $z\approx 10$, \cite{boylan-kolchin_accelerated_2025} additionally points out that the high concentration of high-redshift dark matter halos may naturally support efficient conversion of gas to stars in even lower mass halos through increased gravitational accelerations.
\cite{boylan-kolchin_accelerated_2025} notes that these stars should form in bound, dense clusters similar to globular clusters. 
Our simulations probes many low-mass dark matter halos with high gravitational acceleration, potentially analogous to this model. 
For halos that have experienced initial metal enrichment, several studies investigated the ability of dense clusters to form through mergers \citep[][]{nakazato_merger-driven_2024} or disk collapse \citep[][]{Mayer+24}.
	
Thus, it seems likely that once high surface density regions arise, early systems collapse in a losing battle against feedback processes to generate a high density cluster \citep[e.g.,][]{lahen_mergers_2025}. 
However, resolving the origin, nature, and fate of these dense systems in the broader context of a cosmological picture is less clear.
\cite{Williams+25} investigated the formation of Pop III star clusters in a $\Lambda$CDM cosmological simulation and found that 
in a feedback-free case, stellar densities as high as the observed systems can be produced without any modification to the cosmological paradigm. 
However, that study used a simulation box that did not include the effects of the primordial baryon-dark matter streaming velocity, a $\Lambda$CDM effect that 
should reach highly supersonic values in at least $40\%$ of the Universe by volume. In these regions, it may serve to disrupt the ability of halos to host high density gas and stellar systems \citep[e.g.,][]{Naoz+14,Chiou+18,Williams+23, Hirano+23, chen_supersonic_2025}.
Without accounting for the effects of the streaming velocity, it is possible for simulations in a $\Lambda$CDM framework could overpredict the abundance and typical density of high-redshift systems.

The aims of this study are to expand on \cite{Williams+25} in two ways. Firstly, to provide a more comprehensive picture of star cluster evolution by including the effects of the streaming velocity. 
As a part of this investigation, we explore the impact of the streaming velocity on structure identification in the simulation and the dynamical state of the stellar systems, using the baryon-focused algorithm introduced there.
Secondly, although {\it JWST} has not conclusively observed Pop III systems or star clusters at $z>11$ at the time of this writing \citep[but see][for interesting candidates]{fujimoto_glimpse_2025,cullen_ultraviolet_2024,maiolino_jwst-jades_2024},  we explore the possible detectability of such systems. 
Because the emission properties of Pop III systems remain uncertain, we estimate systems' UV luminosity through a simple analytic relation and explore a range of parameters that represent likely emission scenarios.

This paper is organized as follows.
Our simulations and structure identification methodology are presented in \S~\ref{sec:methods}. 
We present the results in \S~\ref{sec:results}. 
In \S~\ref{subsec:identification}, we explore the stream velocity's role in accurately detecting observable stellar clusters. 
We present the surface densities of our star clusters with and without the stream velocity in \S~\ref{subsec:stellardensities} and compare to observed {\it JWST} high redshift clusters. 
We provide an estimate of the {\it JWST} detectability of these Pop III star clusters in \S~\ref{subsec:detectability}. 
Finally, we explore the environment and dynamical state of star clusters in \S~\ref{subsec:environment} and \S~\ref{subsec:dynamicalstate}, respectively. 
A discussion and of the work is given in \S~\ref{sec:summary}. For this simulation suite, we use a $\Lambda$CDM cosmology whose parameters are: $\Omega_{\Lambda}=0.73, \Omega_{\rm m}= 0.27, \Omega_{\rm b} = 0.044, \sigma_{8} = 1.7$, and $h=0.71$. All magnitudes are calculated using the AB magnitude system \citep[][]{Oke-74}. 

\section{Methods}
\label{sec:methods}
\subsection{Simulation Suite}

We use two simulation boxes, each with size (2.5 Mpc)$^3$, run from $z=200$ to $z=12$, from the Supersonic Project simulation suite \citep[][]{Lake+23b,Lake+24a,Williams+24, Williams+25}. 
The boxes differ in their initial conditions by the choice of $v_{\rm bc}$, the value of the stream velocity in the region.
One box has ``no stream velocity,” corresponding to a fluctuation where the relative velocity between the dark matter and baryons is zero, and the other is a high stream velocity box. 
The value in the high stream velocity box is chosen to correspond to a region with $v_{\rm bc}=2 \sigma_{\rm bc},$ where $\sigma_{\rm bc}$ is the r.m.s. value of the streaming velocity (6.4 km s $^{-1}$ at $z=200$).                                                                                                                                                  
The simulations were run using the {\tt AREPO} code with a resolution of (768)$^3$ DM particles and (768)$^3$ gas elements, and initial conditions were generated using a modified version of CMBfast \citep[][]{1996ApJ...469..437S}, including the first-order scale dependent temperature fluctuations and the second-order streaming velocity \citep[][]{NB05}. 
The boxes were chosen to mimic high-density regions (such as the regions that form galaxy clusters) with increased early structure formation, obtained by setting $\sigma_8=1.7$. 
This choice enhances the statistical power of our box. 
As we discuss in \S~\ref{subsec:detectability}, we correct our counts for this overestimate using the mass function of \citet[][]{ST+02}.

The simulation suite includes nonequilibrium molecular hydrogen chemistry, cooling, and star formation, but no feedback processes (radiative or supernova). 
The chemistry and cooling is implemented using the GRACKLE library \citep[][]{smith_grackle_2017, Chiaki+19}, including HD and H2 cooling and chemistry for the following primordial species: (e$^-$, H, H$^+$, He, He$^+$, He$^{++}$, H$^-$, H$_2$, H$_2^+$, D, D$^+$, HD, HeH$^+$, D$^-$, and HD$^+$).

Star formation occurs in gas cells that have exceeded the Jeans mass on the scale of the Voronoi mesh cell. 
Gas cells are converted to stars following the prescription described in \cite{Marinacci+19} through a stochastic process.
Gas cells collapse on the free-fall timescale ($t_{\rm dyn} = \sqrt{3\pi/(32G \rho_{\rm g})}$) and are converted to collisionless star particles with the mass of the gas cell that collapsed. 
Simulation snapshots are saved at $z=15,14,13,$ and $12$, with structures identified through post processing algorithms, described below. 
The results presented in this work are compiled from the snapshot at $z=12$, unless otherwise stated.

\subsection{Identification of star clusters}
	\begin{table*}[]
\centering
        \begin{tabular}{|l|l|l|l|}
            \hline
            \ttfamily {\bf Category} & \ttfamily {\bf Description} & \ttfamily {\bf Primary Particle(s)} & \ttfamily {\bf Secondary Particle(s)}\\                
            \hline 
            \hline
            \multirow{3}{*}{\bf DM-centric} & \multicolumn{1}{l}{DM primary $^\dagger$ (standard)} & \multicolumn{1}{|l}{Dark Matter} & \multicolumn{1}{|l|}{Stars \& Gas} \\\cline{2-4}
                                & \multicolumn{1}{l}{Dark Matter \& baryons} & \multicolumn{1}{|l}{Dark Matter, Stars,} & \multicolumn{1}{|l|}{None} \\
                                & \multicolumn{1}{l}{primary} & \multicolumn{1}{|l}{and Gas} & \multicolumn{1}{|l|}{ } \\\cline{2-4}
                                \hline
                \multirow{2}{*}{\bf Baryon-centric}& \multicolumn{1}{l}{Stars \& gas primary$^\dagger$} &\multicolumn{1}{|l}{Stars and Gas} & \multicolumn{1}{|l|}{None} \\\cline{2-4}
                                 & \multicolumn{1}{l}{Star primary} &\multicolumn{1}{|l}{Stars} & \multicolumn{1}{|l|}{None} \\\hline
        \end{tabular}
    \caption{Summary of structure finding algorithms used in this work. $^\dagger$ indicates the algorithm used as the standard for the category in the analysis, where only the difference between the DM-centric and the baryon-centric methods is considered. In other words, when we refer to ``DM-centric" and ``baryon-centric" algorithms in those sections, we mean the DM primary and the stars\& gas primary runs, respectively. } 
    \label{tab:FOFruns}
\end{table*}

In order to identify  star clusters that could be ``detected" electromagnetically in a high-resolution, high-redshift simulation, it is important to track baryonic particles in addition to a standard halo-finding algorithm. 
The {\tt AREPO} default friends-of-friends (FOF) algorithm \citep[][]{Springel2010a} struggles to detect all stellar clumps when targeting the dark matter particles as a primary particle type \citep[][]{Williams+25}.
This occurs because of several factors.
Firstly, halos below the resolution threshold for dark matter may still host resolved star clusters. 
Additionally, merging systems (which are abundant at this early epoch of hierarchical build-up) can cause a halo to be undetected.
Finally, fragmentation, accretion, and merger processes lead to star clusters that are offset from the center of mass of the halo, and thus are missed due to the geometry of the system. 
In this study, we include the stream velocity, which further complicates the task of a dark matter-focused structure finding algorithm.
The stream velocity is known to separate collapsing baryonic and dark matter perturbations, causing offsets between the centers of mass of gas structures and their host halos, which can exceed the scale of the virial radius \citep[e.g.,][]{Lake+24a, Lake+22,Lake+21,Chiou+18,Naoz+14}. 
Furthermore, the stream velocity can affect the density of halos, generating diffuse structures that may be harder to detect \citep[e.g.,][]{Williams+23}.

Thus, for this work, we first briefly compare the catalog of simulated structures as identified with the standard dark matter-focused algorithm with a set of alternative algorithms, first introduced in \cite{Williams+25} and listed in Table~\ref{tab:FOFruns}.
These are two algorithms using dark matter as the primary particle type and two that ignore dark matter and focus on baryonic material. 
The baryonic-focused algorithms are meant to imitate observations with no knowledge of the underlying dark matter field. 
The first of these is an algorithm that targets star particles as the only primary particle type, and the second targets both stars and gas particles. 
When we compare with {\it JWST} observations, we use the baryonic algorithm with stars and gas as the primary particle types. 
This algorithm is less susceptible to fragmentation and resolution errors.
For our dark matter-focused algorithms, we test a ``standard" version with dark matter as the primary particle type and stars and gas particles associated at a later stage. 
Additionally, we target all particle types (dark matter, stars and gas) as the primary type for the final algorithm. 
As shown in \cite{Williams+25}, this algorithm typically traces the dark matter version, but is included for completeness in the comparison. 

For dark matter halos, we enforce a resolution threshold of 300 particles, discarding all objects in the catalog that do not meet this criterion.
For baryonic-detected star clusters and galaxies, we require that systems have at least 100 star particles, with no requirement on the gas cells. 
To ensure that no unbound structures are included (nor unrelaxed systems mistakenly identified as virialized), we tested the robustness of our results in the low-particle-number regime (see Appendix of \citealt{Williams+25} and Appendix~\ref{app.conv} below).
We find that the scatter in virial ratio does not change over an order of magnitude variation in particle number, and furthermore, our dynamical analysis is robust even if the particle resolution threshold is increased by a factor of five.
The gravitational softening length at the end of our simulation is $\sim 1$ pc, and thus we do not include objects whose half mass radius is less than this value as the gravitational interactions will not be resolved. 
The catalogs without the stream velocity are the same as the public arepo-clusters\footnote{ \href{https://www.astro.ucla.edu/~clairewilliams/cluster-catalog}{https://www.astro.ucla.edu/~clairewilliams/cluster-catalog},  DOI \href{https://doi.org/10.5281/zenodo.15392998}{10.5281/zenodo.15392998}} \citep[][]{cluster-catalog} dataset.
Post-processing for the catalogs with the stream velocity was performed using the process-fof\footnote{\href{https://github.com/astro-claire/process-fof.git}{https://github.com/astro-claire/process-fof.git}, DOI \href{https://doi.org/10.5281/zenodo.15701630}{10.5281/zenodo.15701630}} code \citep[][]{process-fof}, the same code used for the catalogs without the stream velocity.

\subsection{Dynamical state}

The dynamical state of these star clusters is interesting because it traces the process of structure formation at these times. 
We test each structure to ensure that the system is gravitationally bound (total energy $E<0$) and exclude those that do not meet this criterion from further analysis. 
We additionally calculate the kinetic and potential energy of each stellar system and test whether it meets our criterion for virialization, which is that the ratio of the potential energy ($U$) to the kinetic energy ($K$) is: 
\begin{equation}
   -2.5<U/K <-1.5. 
\end{equation}
Additionally, in the analysis below, we label bound objects outside the above criterion as super virial:  
\begin{equation}
    -2.5\le U/K, 
\end{equation}
or sub virial:
\begin{equation}
    -1.5\le U/K \le 0.
\end{equation}
To calculate the energies, a summation is made over every star particle in the object. If the boundedness or virialization criterion is not met, we additionally test the summation including all dark matter particles within the maximum radius. 
This process is described in detail in \cite{Williams+25} and a flow chart schematic is reproduced in  Appendix~\ref{App:catalog} to illustrate this methodology (Fig.~\ref{fig:flowchart}). 
We calculate the maximum radius, the half-mass radius (the minimum radius that encloses half of the object's stellar mass), and the virial radius ($R(U/K=2)$), if the object is virialized. 
To associate dark matter halos, we calculate both the total mass of dark matter particles enclosed within the object’s radius and find the nearest halo in the dark matter halo catalog.

\subsection{Semi-analytic model}

	In this work, we seek to estimate the intensity of the potentially observable emission from the structures by {\it JWST}. As the simulation does not explicitly model the emission and radiative transfer of stars and gas, we rely instead on a simple model to produce approximate values. This has the advantage that key parameters that are uncertain (such as the Pop III mass distribution) can be easily varied to produce a range of possible values. 
    In this framework, the UV  emission (as observed redshifted into the infrared by {\it JWST}) is assumed to trace the young, massive stars whose stellar continuum dominates this portion of the spectrum. 
    Thus, we make use of the standard literature relation 

\begin{equation}
    L_{\rm UV,1500} = \dot{M}_* /\mathcal{K}_{\rm UV, 1500},
    \label{eq:uvluminosity}
\end{equation}
where  ${\cal K}_{UV,1500}=1.15 \times 10^{-28}$  $M_\odot$ yr$^{-1}$ / (ergs s$^{-1}$ Hz$^{-1})$ is the constant representing the UV photon production of the newly-formed stellar population, and $\dot{M}_*$ is the star formation rate in units of $M_\odot/$yr.
This relation assumes a \citet{salpeter_luminosity_1955} initial stellar mass function.
We note that the value of ${\cal K}_{UV,1500}$ depends on the underlying stellar population \citep[e.g.,][]{Madau+14,tacchella_redshift-independent_2018}, and if the IMF of the Pop III stars here is top heavy, the value of $\mathcal{K}_{\rm UV, 1500}$ can be as low as ${\cal K}_{UV,1500}=0.3 \times 10^{-28}$  $M_\odot$ yr$^{-1}$ / (ergs s$^{-1}$ Hz$^{-1})$ \citep[e.g.,][]{harikane_comprehensive_2023,zackrisson_spectral_2011}. The simulation does not include dust, and we do not add any dust effects or correction in this semi-analytical framework.

Estimating the star formation rate is also necessary for an appropriate observational comparison.
Using the snapshots available to us in this simulation suite, we determine the number of new stars in each object by searching for the particle IDs that do not appear in the previous snapshot. 
This gives us a total mass of new stars in each object ($M_{\rm *, new})$. 
We note that this is a much more accurate method than that used in \cite{Williams+24}, where the difference in stellar mass was constructed from a merger tree, which did not actually take into account whether or not a star particle had formed since the last snapshot.
From here, we  estimate the timescale over which these stars formed.
A lower limit on the star formation rate is thus provided by assuming these stars formed at a constant rate over the entire duration between the previous snapshot and the snapshot of interest ($\Delta t_{\rm snapshot}$). This gives:
\begin{equation}
    \dot{M}_* \approx \frac{M_{\rm *, new}}{ \Delta t_{\rm snapshot}},
    \label{eq:starformationrate}
\end{equation}
which we use to estimate the luminosity of our simulated star clusters via Eq.~\ref{eq:uvluminosity}. The typical $\Delta t_{\rm snapshot}$ varies with redshift but is on the order of 10 Myr, comparable to the timescale over which the rest frame UV/optical emission used in observations traces star formation.

\section{Results}
\label{sec:results}

\begin{figure*}
    \centering
    \includegraphics[width=0.45\linewidth]{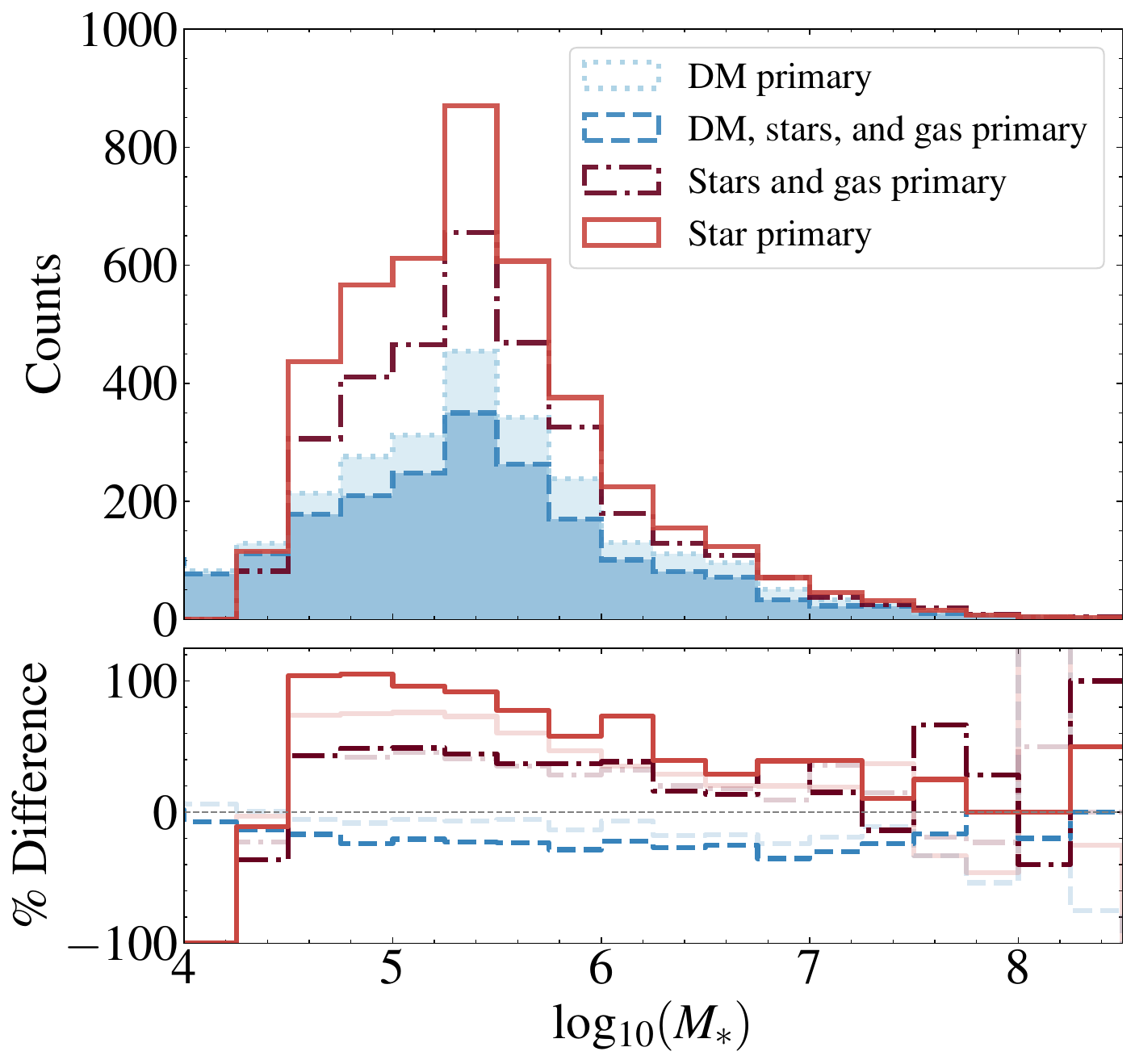}
    \includegraphics[width = 0.45 \linewidth]{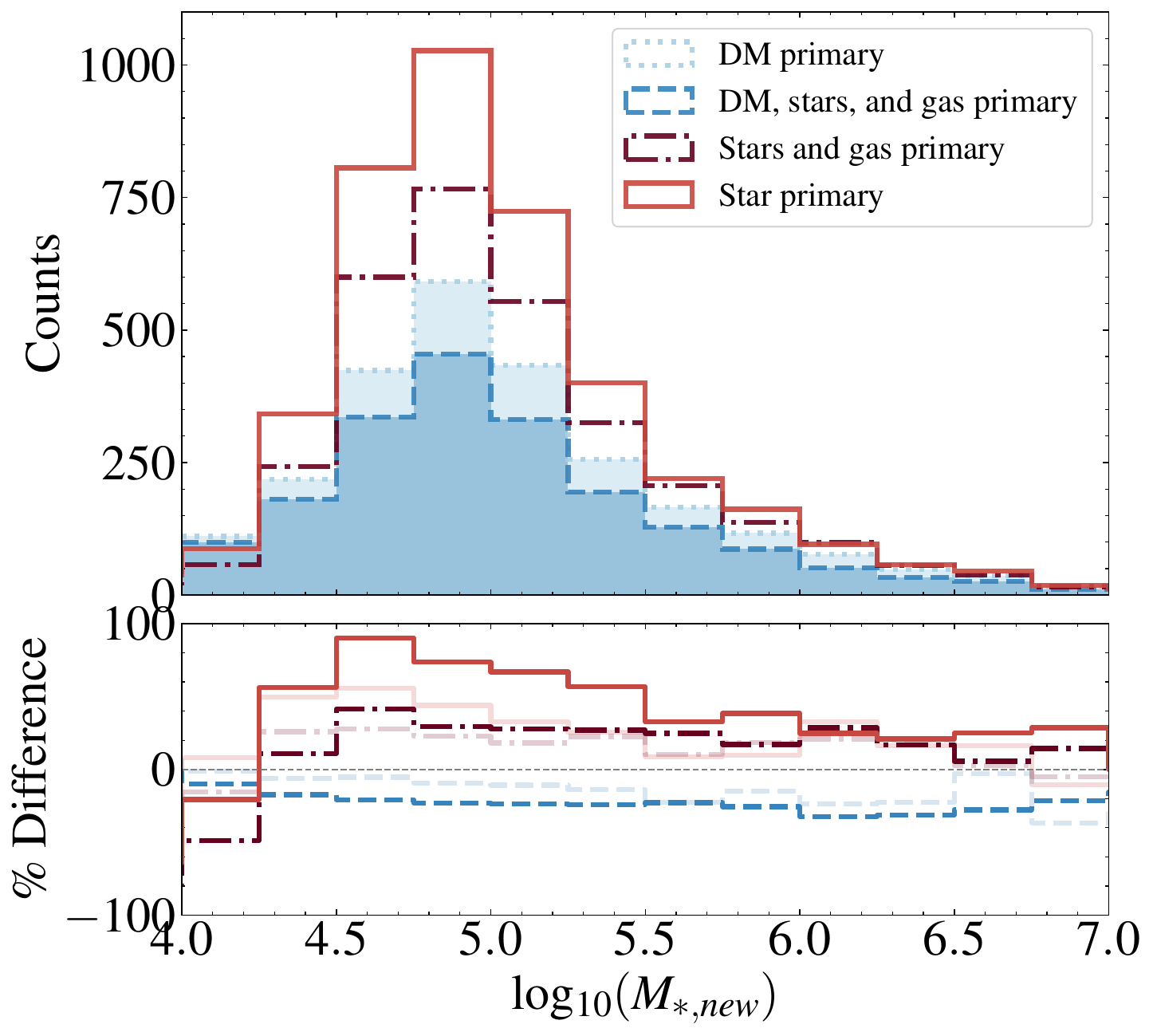}
    \caption{Histograms by stellar mass (left) and newly-formed stellar mass (right) at $z=12$ in the simulation box including the stream velocity. 
    The bottom panel shows the percent difference between the counts in the dark-matter primary and the other algorithms. 
    In the bottom panel, the percent differences observed when the stream velocity is not included (data from \citealt{Williams+25}) are shown as faint lines. 
    The different line colors denote the various structure finding algorithms. Blue runs are structure finding algorithms that include dark matter, and red lines correspond to structure finding algorithms focused on baryonic particles only.
    Pale blue dotted lines are dark matter primary structures. Dark blue dashed lines show the DM, stars, and gas primary.
    Red solid lines denote star primary, and dark red dot-dashed denotes stars $+$ gas primary algorithm (used in the further analysis here). }
    \label{fig:mstar_hist}
\end{figure*}

\subsection{Identification of simulated star clusters} 
\label{subsec:identification}
First, we test the structure finding algorithms for their ability to find star clusters when the effects of the stream velocity are present. 
\cite{Williams+25} already showed that the algorithms disagree when no stream velocity is included. 
Thus, here we seek to understand:
\begin{itemize}
    \item[(a)] Whether or not there is a discrepancy in the number of objects detected between the DM halo catalog and catalogs that use the baryonic particles when the stream velocity is included. 
    \item[(b)] If so, whether this discrepancy is larger for the box with the stream velocity than without the stream velocity. 
\end{itemize}
We address these questions in Fig.~\ref{fig:mstar_hist}.
The top left panel shows histograms of objects in the four catalogs, binned by stellar mass, and the bottom panel shows the percent difference between each algorithm and the standard dark-matter focused version.
The data in the top panel represent the objects identified in the stream velocity box.
It is clear from these counts that indeed, the catalogs which include dark matter and the catalogs which only use baryonic particles are discrepant in their number counts. 
The baryonic catalogs (star primary and stars + gas primary) both typically detect more objects, in agreement with the findings of \cite{Williams+25}.

To address point (b) above, in the bottom panel, which illustrates the percent difference between the two cases, the non-stream velocity data from \cite{Williams+25} is shown with the fainter lines. 
At low masses ($<10^6 M_\odot$), the discrepancy between the star-primary and the dark-matter-primary is higher with the stream velocity than in the no stream velocity case (exceeding 100\%). 
This is the case for every algorithm, but especially for the star-only catalog. 
When gas is included in the algorithm as a primary particle, this discrepancy is reduced. 
This suggests that using gas particles may identify baryonic objects with multiple stellar sub-clumps that are part of one structure linked by a gaseous component. 
This scenario may occur more frequently in the box with streaming included, where halos whose gas has been advected via the stream velocity are now re-accreting this gas. 
However, Fig.~\ref{fig:mstar_hist} may also imply that the gas+stars algorithm is missing some star clusters with little or no gas component.

The right panel of Fig.~\ref{fig:mstar_hist} shows the objects binned 
by newly formed stellar mass ($M_{\rm *,new}$), the parameter which is used to estimate the star formation rate (see Eq.~\ref{eq:starformationrate}). 
Again, discrepancies between the runs are present. 
However, the gas and stars-focused run and the stars-only run converge with each other at high values of $M_{\rm *,new}$, which is the most important regime for observational comparison, as we expect these objects to be brightest. 
Searching using the baryon-centric method underpredicts the low-mass end of the $M_{\rm *,new}$ distribution. 
This is because of the resolution cutoff: only structures with 100 or more star particles are considered, so systems with extremely low stellar masses, and thus low numbers of new stars, are not counted (this can be seen in the left panel as well). 
We note that since many young high-redshift galaxies are dominated by nebular emission lines for young stellar populations, including the gas as well as the stars, the primary algorithm should select for these more easily detectable systems.
For these reasons, we continue the analysis using the gas and stars-focused structure finding algorithm to generate the catalog of star clusters for this work. 

\subsection{Stellar densities}
\label{subsec:stellardensities}

\begin{figure*}
    \centering
    \includegraphics[width=0.7 \textwidth]{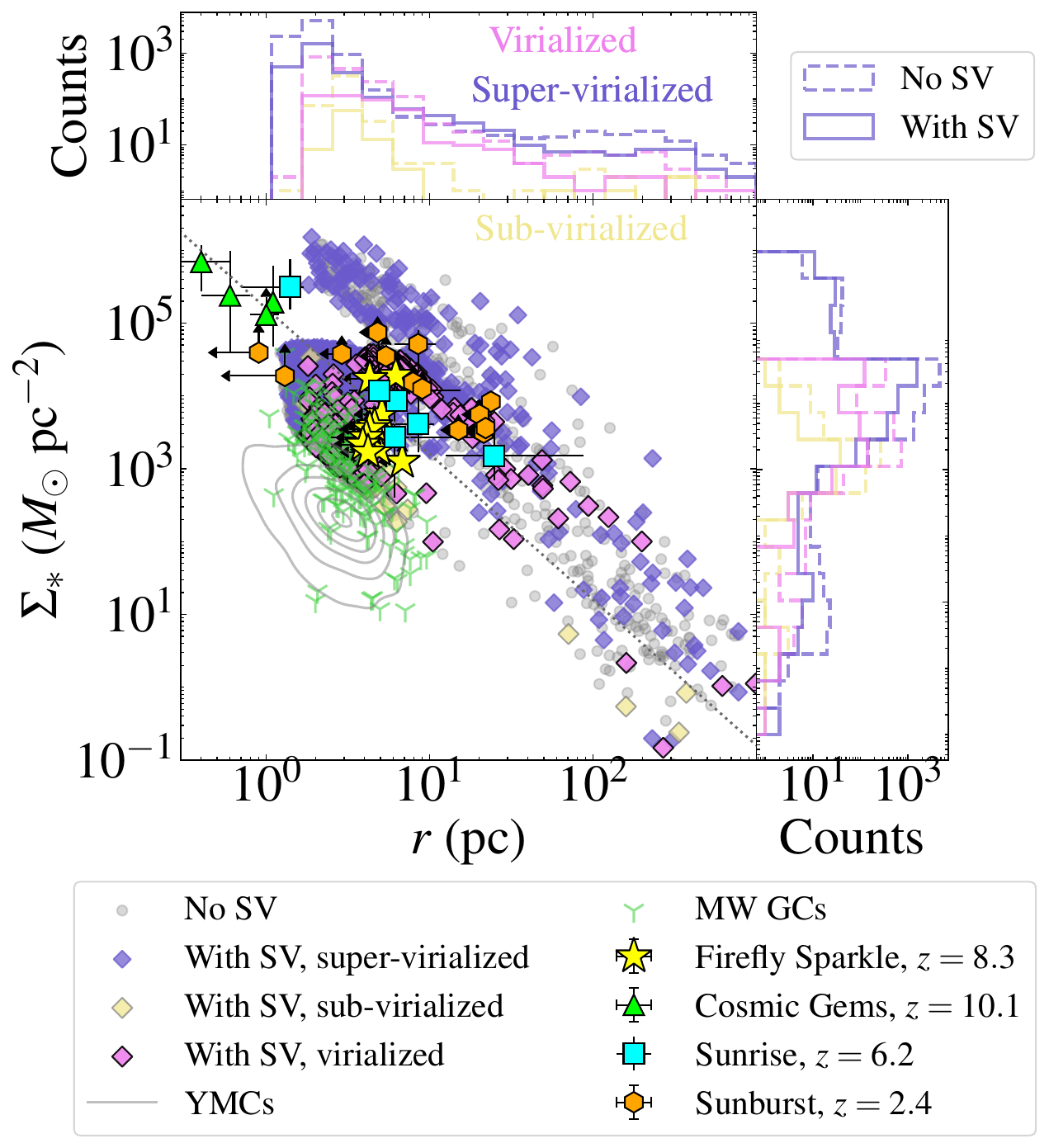}
    \caption{Main panel: stellar mass surface density versus radius for simulated objects at $z=12$ in comparison to literature observations. 
    Purple diamonds show the super-virialized star clusters, yellow diamonds show sub-virial star clusters, and pink diamonds are the virialized star clusters. 
    The no-SV data is shown in grey underneath (as in \citealt{Williams+25}). 
    On the right and top panels, histograms are presented showing the projected distributions of the objects from both main panels along the $y$- and $x$-axes. 
    The colors follow the same catagories as the main panel.
    For these histograms, solid lines correspond to the stream velocity box, and dotted lines correspond to the no-stream velocity box. Overplotted in the main panel are milky way globular clusters  \citep[green Y][]{Gieles+11}, local young massive clusters \citep[YMCs,][]{Brown+14}, and high redshift lensing observations at redshifts between $2-10$ \citep[][]{Adamo+24, Mowla+24, Vanzella+22, Vanzella+23}. The dotted line shows the expected behavior of a uniform density $10^6M_\odot$ star cluster.}
    \label{fig:surface_density}
\end{figure*}

Using the gas and stars (baryonic-focused) algorithm to identify star clusters, we now turn to estimating their densities to compare with observations of star clusters at high redshift. 
In Fig.~\ref{fig:surface_density}, we plot the stellar surface density ($\Sigma_*$) against the half mass radius ($r_{\rm hm}$), for clusters in the box including the stream velocity.
The surface density is estimated assuming a uniform sphere within the half mass radius.
For comparison, the non-stream velocity points are shown underneath as grey points. 
The half-mass radius is used to compare with the size of observed objects, since the diffuse starlight at the outskirts of systems is unlikely to be detected.
Both the stream velocity and no stream velocity simulations result in systems spanning a similar range in half-mass radius. 
This can be seen in the upper histogram panel, which shows the distribution of simulated points on the $x$-axis.
In the histograms, dashed distributions show the no-stream velocity box, and the solid distributions correspond to the box with the stream velocity.
The objects are split by their virialization state, which will be discussed further in \S~\ref{subsec:dynamicalstate}.
From the right histogram panel, showing the stellar surface density (projection of the simulated points along the $y-$axis), it is clear that  star clusters achieve similarly high densities,  up to $\sim 10^6M_\odot$ pc$^{-2}$, in both cases, although the box with stream velocity contains fewer clusters overall.
We overplot the observed {\it JWST} clusters from \cite[][]{Adamo+24, Mowla+24, vanzella_early_2022, Vanzella+23}. 
The Figure suggests that $\Lambda$CDM clusters can reach the observed high densities of {\it JWST} lensed observations (for the region of the parameter space that is resolved by the cosmological boxes considered here) even when there is a strong stream velocity in the region. 
For comparison, the green Y-shaped points correspond to Milky Way globular clusters \citep[][]{Gieles+11}. 
The gray contours show the distribution of young massive clusters \citep[YMCs,][]{Brown+14}. 
We note that the simulation also produces many objects with radii $<10$ pc and surface densities between $10^2-10^3\,M_\odot$ pc$^{-2}$, similar to globular clusters and the higher density YMCs, as well as a tail of large, low density clumps which may represent larger, extended protogalaxies. 
Although the inclusion of feedback may disrupt the formation of such lower-density systems (see discussion), if they are present they are likely more challenging to detect due to reduced surface brightness. 
This leads us to an investigation of luminosity and detectability in the following section.

\begin{figure}
    \centering
    \includegraphics[width=0.95\linewidth]{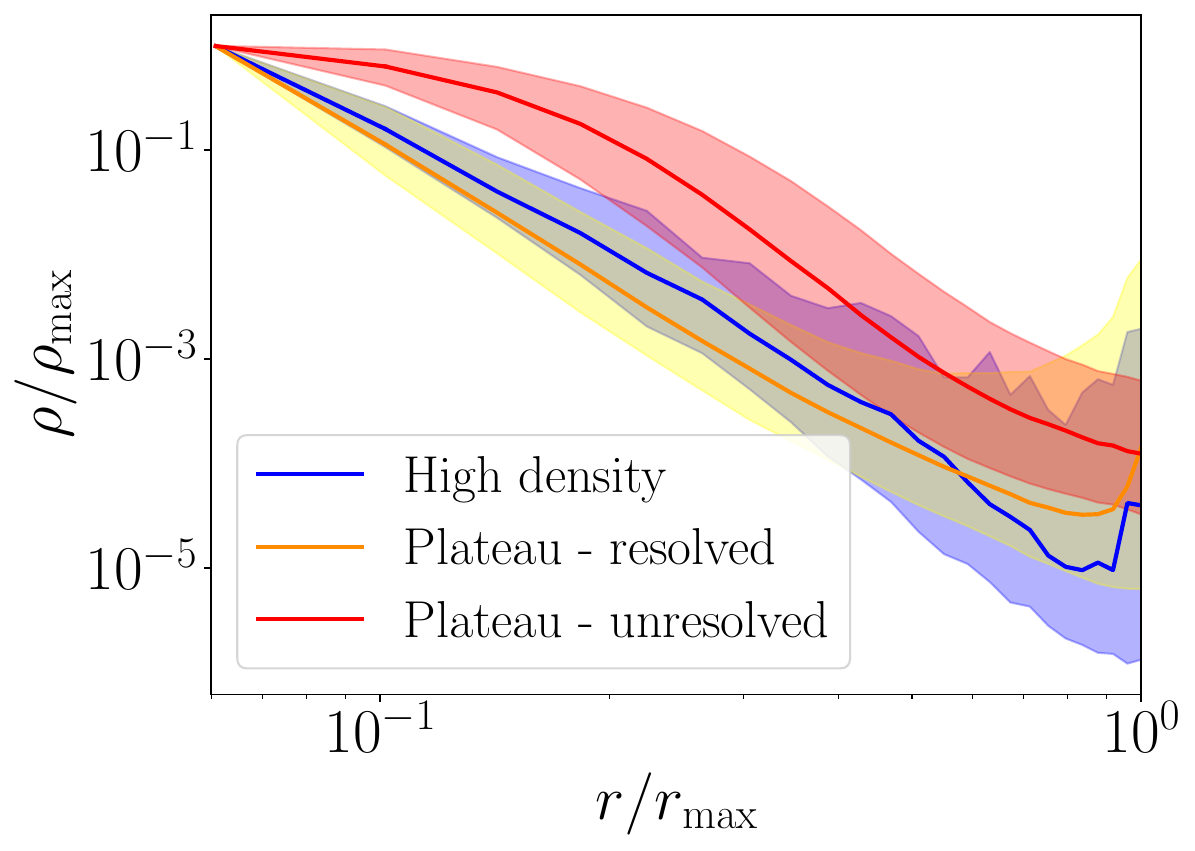}
    \caption{Median normalized density profiles. 
    The blue curve shows the median profile of the subset of objects with extremely high surface density, $M_*>10^{5}M_\odot$ pc$^{-2}$ (98 systems). 
    The orange and red curves show objects the ``plateau" discussed in \S~\ref{subsec:stellardensities}, meeting the two criteria that $M_*<10^{5}M_\odot$ pc$^{-2}$ and $r_{\rm h}<20$pc (10563 objects).
    All objects in the high-density peak are fully resolved over the range of radii depicted. 
    However, for some low-mass objects in the plateau, the inner radii are not resolved within the softening length of the simulation. 
    Thus, we divide the plateau into resolved and unresolved systems, for which the gravitational softening length is greater than $10\%$ of the cluster's maximum radius. 
    The medians are smoothed using a Savitzky-Golay filter with a window length of 5 and a 2nd order polynomial.
    The shaded regions show the 16th-84th percentile bands of the distribution. }
    \label{fig:densityprofile}
\end{figure}

Finally, we briefly investigate the triangular gap seen in Fig.~\ref{fig:surface_density} around $10^5M_\odot$ pc$^{-2}$ towards small radii. 
%
We suggest that this gap originates from numerical effects, rather than physical processes. 
In particular, some objects with very low mass become close to the simulation resolution limit in size.  
We investigate these effects in Fig.~\ref{fig:densityprofile}, where the normalized median density profiles of a subset of our systems are shown.
We split our star clusters into  high density ($M_*>10^{5}\,M_\odot$ pc$^{-2}$, systems  above the triangular gap) and plateau ($M_*<10^{5}\,M_\odot$ pc$^{-2}$ and $r_{\rm h}<20$ pc, systems below  the triangular gap). 
High density systems are plotted in blue in Fig.~\ref{fig:densityprofile}. 
For these systems, the softening length is well under 10\% of their maximum radius. 
We thus consider the dynamical state and density of these systems to be ``resolved." 
This gives us confidence that the high densities reached are robust and not due to numerical effects. 
However, for the ``plateau" systems, we find that $\sim60\%$ have the simulation gravitational softening length greater than or equal to $10\%$ of the maximum radius. 
We thus consider the inner star clusters of these systems to be dynamically ``unresolved" in their inner core.\footnote{
We note that in this case, ``unresolved" refers to the dynamical distribution within the inner core, rather than the simulation resolution of 100 star particles--a criterion which is met by all objects in the catalog. 
The robustness of our overall cutoff is discussed in the Appendix~\ref{app.conv}. }

The median density profiles for the resolved and unresolved systems in the plateau are shown in yellow and red in Fig.~\ref{fig:densityprofile}, respectively. 
Indeed, for the ``unresolved" systems, a cored profile shape is clearly seen to a radius $> 0.1\,r_{\rm max}$, while the resolved plateau systems follow a similar cuspy distribution to the high density systems. 
The Figure demonstrates that for a subset of systems, the resolution-driven core extends the half mass radius, artificially decreasing the measured surface density in the simulation. 
Thus, it seems that the lack of star clusters in the triangular region of Fig.~\ref{fig:surface_density} is due to the gravitational softening length, an effect of the simulation's finite resolution. 

We additionally speculate that a dynamical simulation effect plays a role in driving all low mass clusters to a more extended core distribution than expected: the difference in particle mass between the star and dark matter particles. 
The dark matter particles are roughly six times more massive than the star particles (which inherit mass from the simulation's Voronoi gas cells). 
In nuclear clusters, relaxation effects from two particle populations with unequal masses can drive the lower-mass particle to form a core profile \citep[e.g.,][]{rom_2024_dynamics,rom_2025_segregation}. 
It seems that this numerical effect drives the star particles in the central regions of our simulated objects to follow a core profile. 
For low masses, this core profile affects the half-mass radius, whereas for high mass objects, the core effect is on a smaller scale than the half-mass radius.
For comparison, the dashed gray line in Fig.~\ref{fig:surface_density} shows the expected $\Sigma_*-r$ relation for a $10^6M_\odot$ object with uniform density.
This mass intersects the triangular, resolution-driven gap, and gives a rough estimate of the mass where dense clusters may still form in a better resolved simulation.
Because of the effects discussed above, our analysis may actually underestimate the number of high density systems in the simulation box.
However, we note that the inclusion of feedback effects can modulate the central surface density, especially for low mass systems--this is discussed further below.

\subsection{Detectability of Pop III star clusters}
\label{subsec:detectability}
\begin{figure*}
    \centering
    \includegraphics[width=0.48\linewidth]{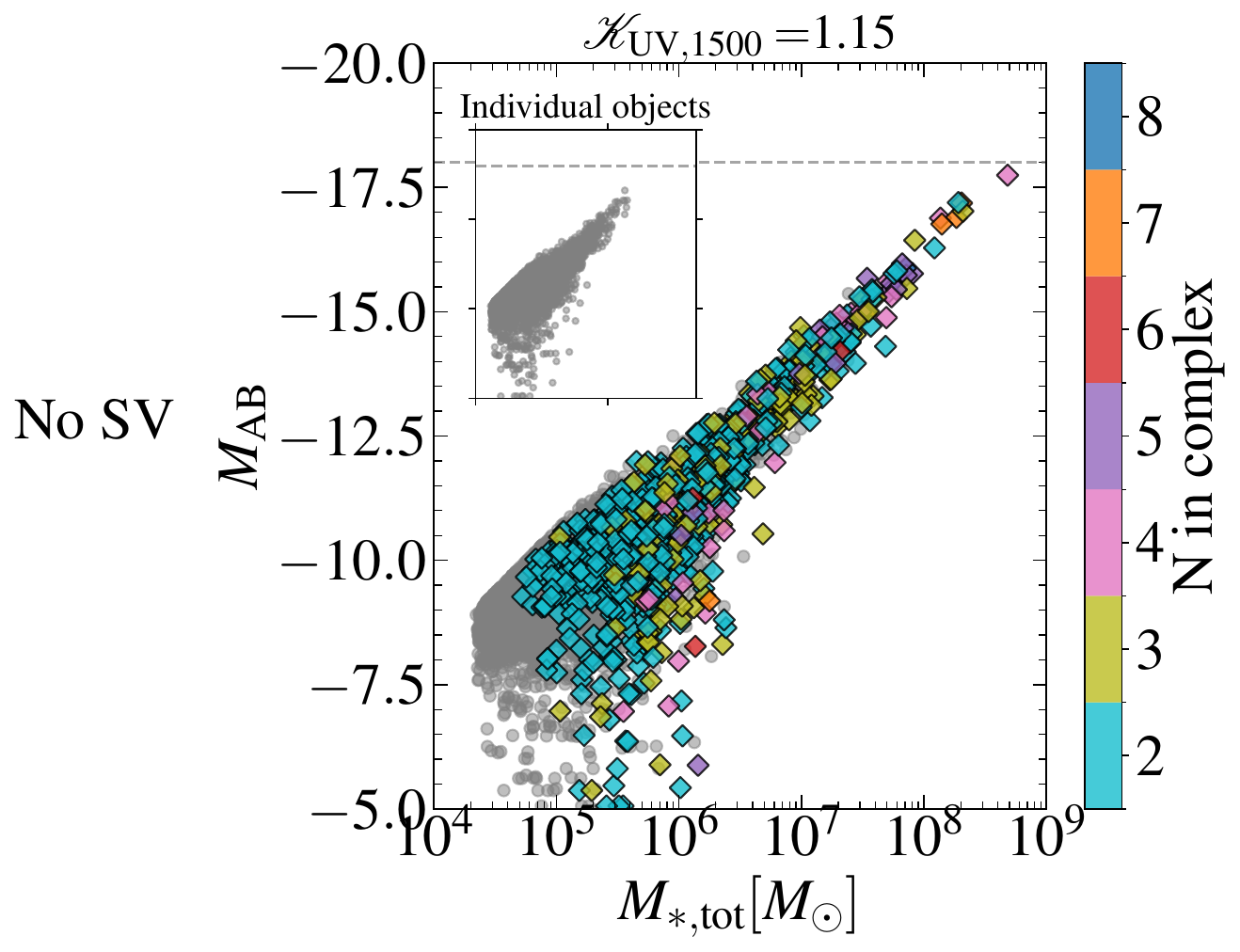}
    \includegraphics[width=0.4\linewidth]{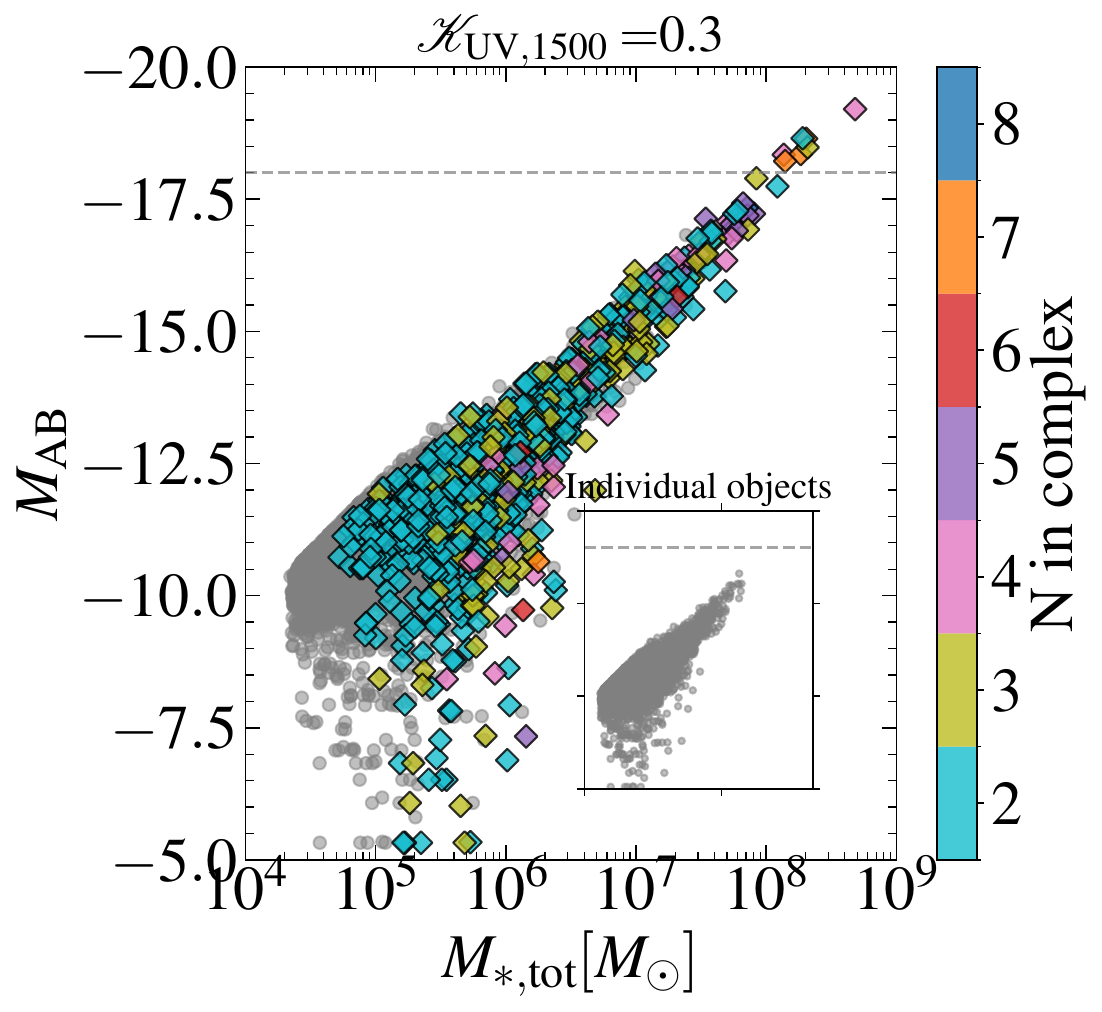}
    \includegraphics[width=0.48\linewidth]{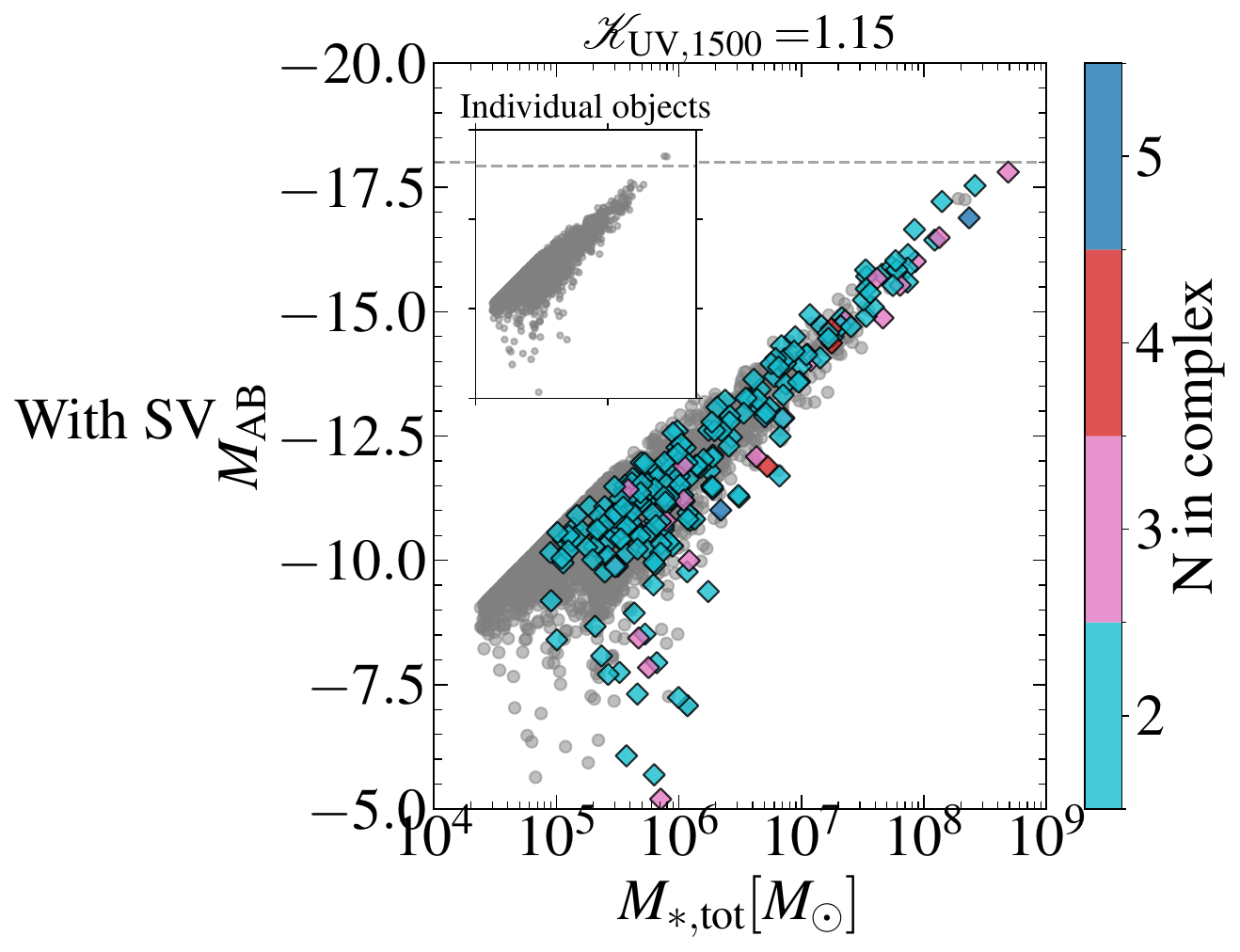}
    \includegraphics[width=0.4\linewidth]{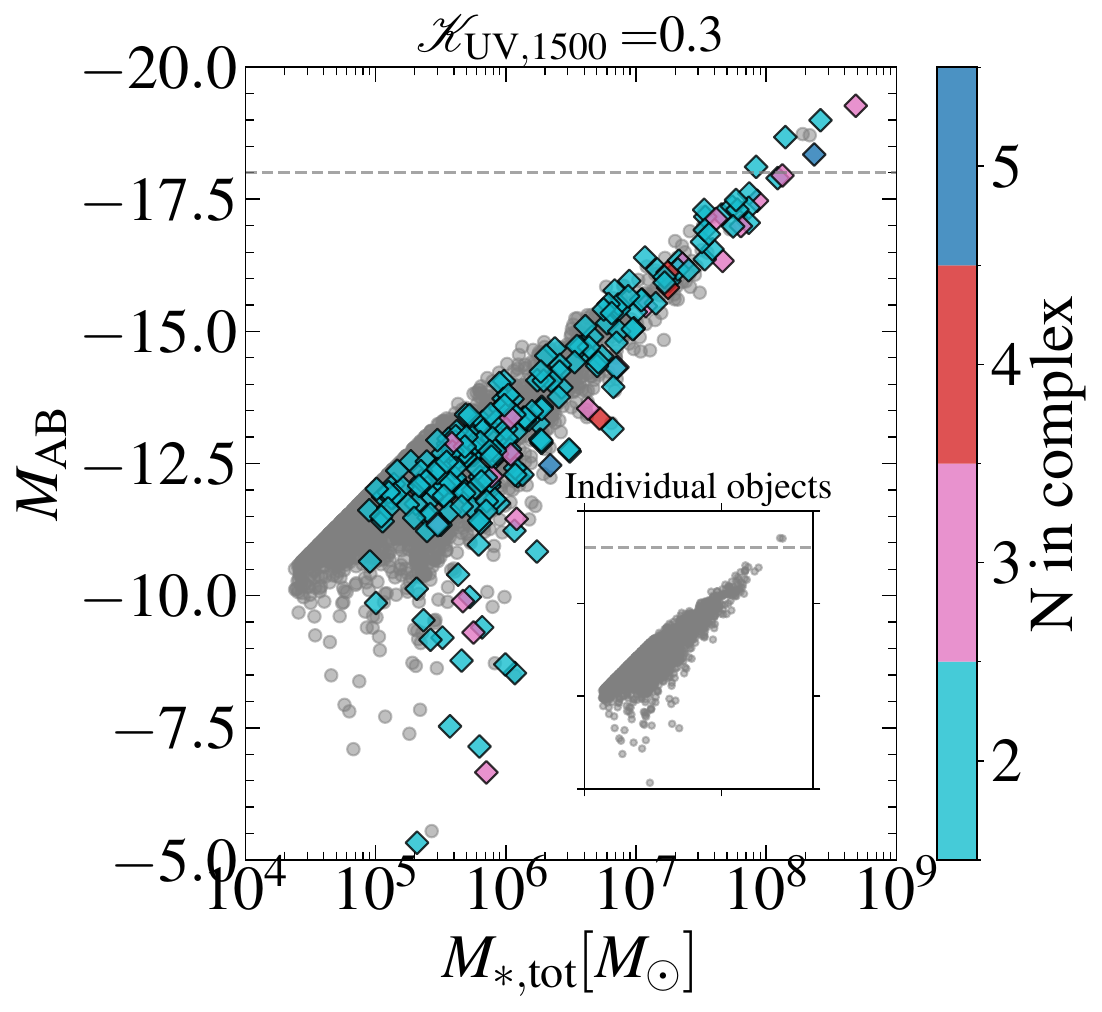}
    
    \caption{Absolute UV magnitude versus stellar mass for clumps and individual objects at $z=12$. The top row is without stream velocity, and the bottom row includes stream velocity. The color bar shows the number of individual star clusters in each of the clumps.
    To guide the eye, a dashed gray horizontal line has been added at $M_{AB} =-18$ in both the main and the inset panels. The right and left columns show the same analysis with differing conversion factors $\mathcal{K}_{\rm UV, 1500}$, in units of $ 10^{-28} M_\odot $ $\text{ yr}^{-1}$$/(\text{ ergs s}^{-1} \text{ Hz}^{-1})$. On the left is the standard value of $1.15$, where as the right represents the increased luminosity of a top heavy IMF.  
    }
    \label{fig:mAB_stellarmass}
\end{figure*}

\begin{figure*}
    \centering
    \includegraphics[width=0.7\linewidth]{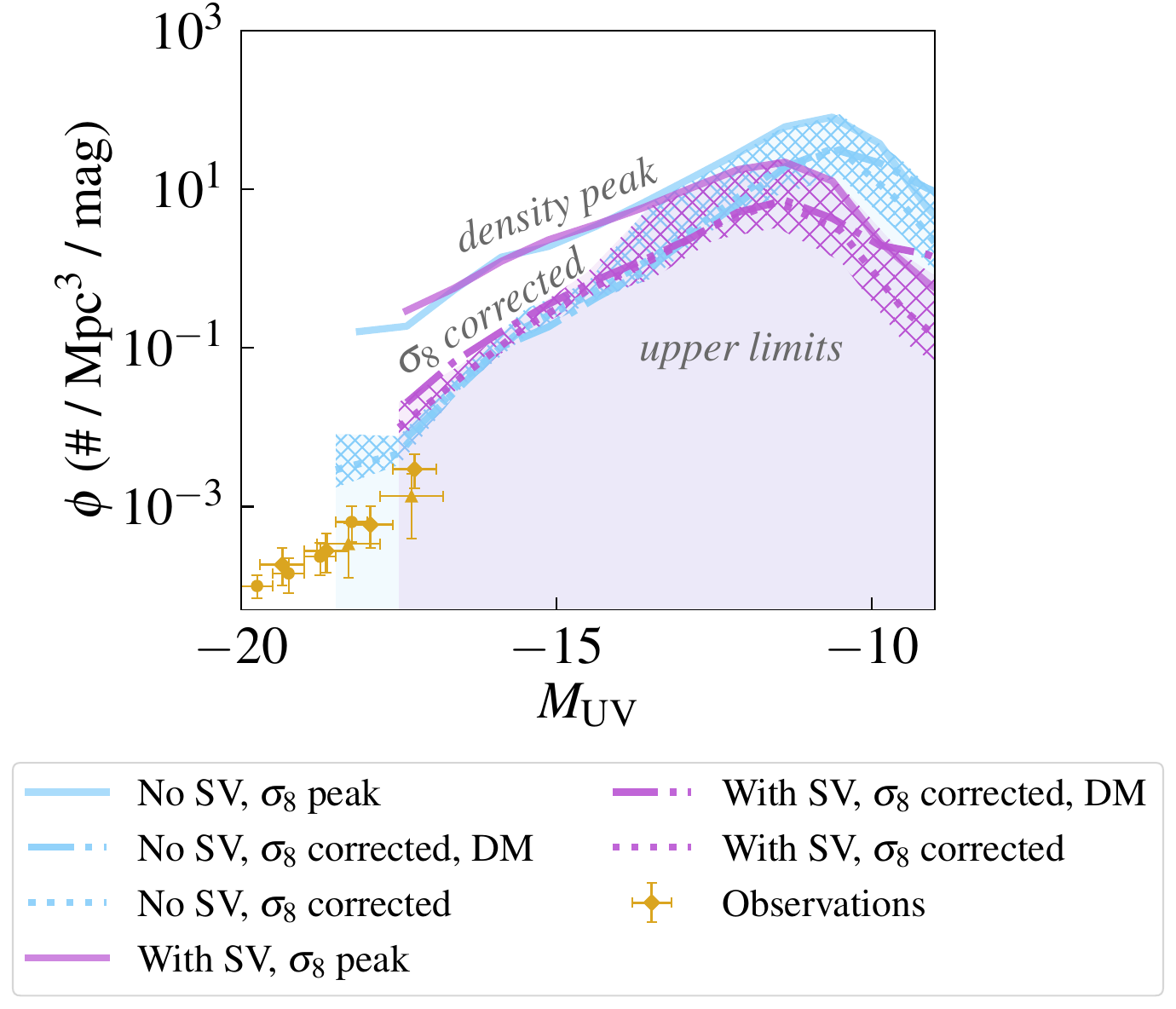}
    \caption{ {\it JWST} $z=12$ UV luminosity function without the stream velocity and with the stream velocity. Counts are determined by the stars+gas primary method, and clumps of several star clusters below the {\it JWST} resolution limit are counted as one object, as in Fig.~\ref{fig:mAB_stellarmass}. 
    The solid lines show the uncorrected luminosity function from our cosmological box, which represents a significant overdensity. 
    The dot dashed and dotted lines have been computed from the baryonic and dark matter catalogs, respectively, including a correction for the increased value of $\sigma_8$ to achieve a luminosity function corresponding to an average region. 
    The hatched regions show  one standard deviation above and below the dotted line due to the spread of dark matter masses in each bin used in the correction factor. 
    We require at least five objects in each magnitude bin. Because feedback is not included, these calculations represent upper limits. The yellow error bars show observational data from: \cite{PerezGonzalez+23, Leung+23uvlf, donnan_jwst_2024}. }
    \label{fig:UVLF}
\end{figure*}

We now seek to provide an estimate of the luminosity of star clusters considered in this work in order to ascertain their detectability by {\it JWST}. 
Because our simulation suite only tracks collisionless ``star particles," representing groups of stars, and we do not trace the metallicity and radiation of those stars, we use the relationship between star formation rate and $1500\AA$ luminosity to provide a useful estimate the brightness of our simulated systems. 
As described in \S~\ref{sec:methods}, we use the star formation rate, estimated based on the number of new stars between the $n$-th simulation snapshot and the $(n-1)$-th simulation snapshot (see Eq.~\ref{eq:starformationrate}). 

\paragraph{Star Clusters}
Because the vast majority of these star clusters are likely to lie below the resolution limit of {\it JWST} (see for example, their radii on the $x-$axis of Fig.~\ref{fig:surface_density}), it is also useful to consider whether clusters of these objects could be detectable in combination.  
For all the star clusters in our simulation boxes whose size is below the {\it JWST} resolution limit, we search for other star clusters within a sphere of radius equal to the effective resolution of {\it JWST} at the redshift of the simulation snapshot (assuming the minimum resolved scale is $0.2$ arcsec). 
In the left column of Fig.~\ref{fig:mAB_stellarmass}, we plot the absolute AB magnitude at $1500 \AA$ in the source frame versus the stellar mass for the complexes and isolated individual objects. 
The gray points represent single objects, while the colored diamonds represent the combined mass and luminosity of multiple star cluster complexes.
The color bar denotes the number of objects that make up the complex. 
The inset panel highlights the individual objects that are not part of any clump. 
Clumps of a few star clusters separated by a distance smaller than the telescope's resolution
are the dominant population in the bright regime.
The brightest isolated objects are typically several magnitudes fainter than the brightest complexes.
However, upon individually investigating a subset of the systems in this regime, we find that the majority of the complexes have one bright star cluster that is at least an order of magnitude more luminous than its neighbors. 
Thus, the contribution of the fainter or lower-mass neighbors may be small, depending on the system at hand. 
The bottom row displays systems with stream velocity, while the top row illustrates the case without stream velocity. When stream velocity is present, the maximum number of objects in a single cluster is five, compared to eight in the case without stream velocity. We discuss this further below.

In the left panel,  a handful of systems reaches almost $M_{AB} =-18$, a magnitude range comparable with the faintest systems detected by {\it JWST} at $z\sim 12$ \citep[e.g.,][]{donnan_jwst_2024}.
These are the most massive complexes in the entire box, with stellar masses approaching $10^9 M_\odot$. 
These structures, consisting of several sub-clumps around one larger host cluster, would most likely be classified as low-mass galaxies. 
The presence of multiple sub-clusters, 
whose luminosity (and therefore their star formation rate) is reduced at $z=12$ compared to previous redshifts, suggests that one star-forming region in the area may seed the formation environment of low-mass galaxy with metals. 
In other words, low-SFR sub-clumps at $z=12$ may represent systems that formed earlier, providing metal enrichment prior to this simulation snapshot. 
Thus, these will not appear as pristine Pop III galaxies. 
We suggest that the high mass structures in our box are thus roughly consistent with observed galaxies at $z\sim 11-13$ observed in JWST surveys with modest stellar masses around $10^8M_\odot$ \citep[e.g., ][]{Donnan+23b}.
However, our inferred magnitudes are slightly fainter, $-17 \gtrsim M_{\rm UV}\gtrsim-18$, rather than  $-18\gtrsim M_{\rm UV}\gtrsim-19.6$ quoted by \cite{Donnan+23b} for similar masses.

This analysis assumes $\mathcal{K}_{\rm UV, 1500} = 1.15 \times 10^{-28}M_\odot $ $\text{ yr}^{-1}$$/(\text{ ergs s}^{-1} \text{ Hz}^{-1})$. The value of this conversion factor reflects assumptions about the underlying stellar population. 
We also assume that the star formation occurred at a constant rate in the time between simulation snapshots. 
However, observing the stellar population in the early phase of a burst of star formation, as may be very prevalent in the early Universe \citep[e.g.,][]{sun_seen_2023}, can provide a large boost in UV luminosity. 
Additionally, for the Pop III star clusters considered here
massive and extremely massive stars may be present in much greater abundance given the low metallicity of the natal gas. 
The investigation of specific models of Pop III or detailed star-formation burst scenarios is beyond the scope of this work. 
However, in order to explore a scenario where the luminosity output for the stellar population is enhanced due to any of these factors while remaining agnostic to specific mechanism, we additionally plot the same relation with $\mathcal{K}_{\rm UV, 1500} = 0.3\times 10^{-28}M_\odot $ $\text{ yr}^{-1}$$/(\text{ ergs s}^{-1} \text{ Hz}^{-1})$ in the right column of Fig.~\ref{fig:mAB_stellarmass}. 
This corresponds to the Top Heavy Pop III model of \cite{zackrisson_spectral_2011} \citep[as in][]{harikane_comprehensive_2023}. 
This significantly increases the apparent magnitude of clusters. 
The brightest complexes in both boxes reach $M_{AB}=-18$ to $M_{AB}=-19$, more consistent with \cite{Donnan+23b}. 
Additionally, lower mass (i.e., $10^7M_\odot$) clusters have $M_{\rm AB}\sim -17$, possibly suggesting their detectability in a lensing configuration.

Both stream-velocity and no-stream-velocity objects produce a similar distribution of densities and luminosities at high SFR, making it challenging to determine the formation channel of an observed object. However, as indicated by the color bar, a significant difference between the two scenarios is the number of sub-clumps that comprise a given structure. 
When stream velocity is considered, it is quite rare for more than two objects to be co-located, with a maximum of five objects found within one complex. 
In contrast, in the absence of stream velocity, many systems contain three or more components, with a maximum of up to eight. 
In the next section (\S~\ref{subsec:environment}), we investigate the environment of these clusters in greater detail.  

\paragraph{The UV Luminosity Function}
When counting objects and their star formation based on dark matter halos, \cite{Williams+24} found that the faint end of the UV luminosity function (UVLF) was boosted for the stream velocity due to the strong bursts occurring in a single clump. 
However, the methodology used here, which employs baryonic structure to mimic observational probes, can provide a more accurate estimate of the UVLF because it searches for individual objects, regardless of the underlying halo distribution, which is hidden from observational techniques. 
In Fig.~\ref{fig:UVLF}, we plot the UVLF in our simulation boxes based on the baryonic-primary method of structure detection (solid lines). 
To estimate the UVLF most appropriately for observations, we count clumps of star clusters below the {\it JWST} resolution limit as one object (as in Fig.~\ref{fig:mAB_stellarmass}).
Importantly, this is not the average UVLF in the Universe but rather the distribution in a high-density peak (corresponding to our simulation box, which has $\sigma_8=1.7$). 
For this reason, the value of the curves on the y-axis is significantly increased with respect to an expected average UVLF on large scales. 

Thus, in order to achieve the relevant comparisont to observations, we approximate a correction to the high $\sigma_8$ value used in this simulation. 
We use the ratio between the \cite{ST+02} halo mass function with $\sigma_8 = 1.7$ and $\sigma_8 = 0.8$ to correct the number counts at each magnitude. 
This ratio is a function of halo mass, with a larger correction at lower masses (see Fig.~6 of \citealt[]{Williams+24} for example). 
Thus, we use the correction factor corresponding to the average halo mass in each magnitude bin.
The dotted and dot-dashed lines show the resulting corrected UVLF for the baryonic and dark matter catalogs, respectively. 
For the baryonic catalog, we correct using the nearest dark matter halo to each stellar object, and as seen in the Figure, this has good agreement with the results from the DM catalog. 
We note that these curves likely offer a more appropriate observational comparison, but there is some intrinsic scatter given the range of halo masses contributing to each bin. 
One standard deviation is shown by the hatched regions to illustrate the degree of scatter. 
The difference between the solid and dotted lines can be understood as the cosmic variance effect of an average versus high density patch of the universe, while the difference between the blue and the purple curves is the contribution of the stream velocity.
We note that even these corrected curves should be considered upper limits, due to the lack of feedback in our simulations. 
The predicted UVLF lies slightly above the {\it JWST} observations depicted in the plot.

The blue lines, corresponding to the box without the stream velocity are heightened for faint objects. 
For brighter objects, the reduced suppression of structure and the burst of concentrated star formation described in \cite{Williams+24} only serve to allow the stream velocity systems to reach similar counts as the non-stream velocity systems, rather than significantly surpass the blue curve.
Thus, at the bright end of our curve, we see that the difference between the stream velocity and non-stream velocity runs is not distinguishable given the scatter and the size of the cosmic variance effect. 
However, at low masses, the difference between the blue and purple curves is much larger than the difference between the solid and dotted lines. 
This suggests that the stream velocity effects will be present and discernible in the UVLF in this extremely faint regime, around $M_{UV}>-12$, which is challenging with the current facilities. 
Future observatories that could access this regime would see around 100 times more small clusters.

\subsection{Environmental Effects}
\label{subsec:environment}

\begin{figure}
    \centering
    \includegraphics[width=0.9\linewidth]{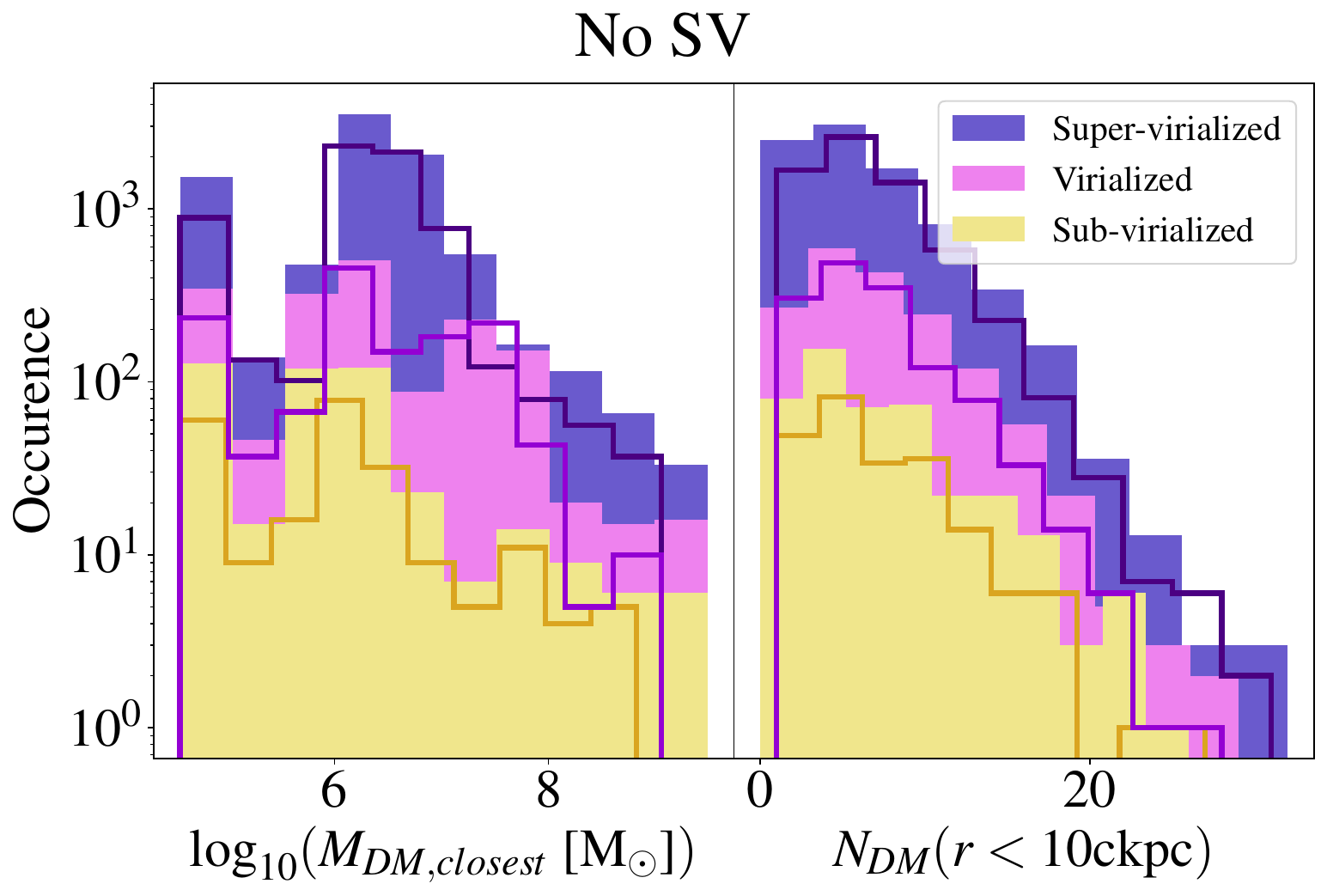}
    \includegraphics[width=0.9\linewidth]{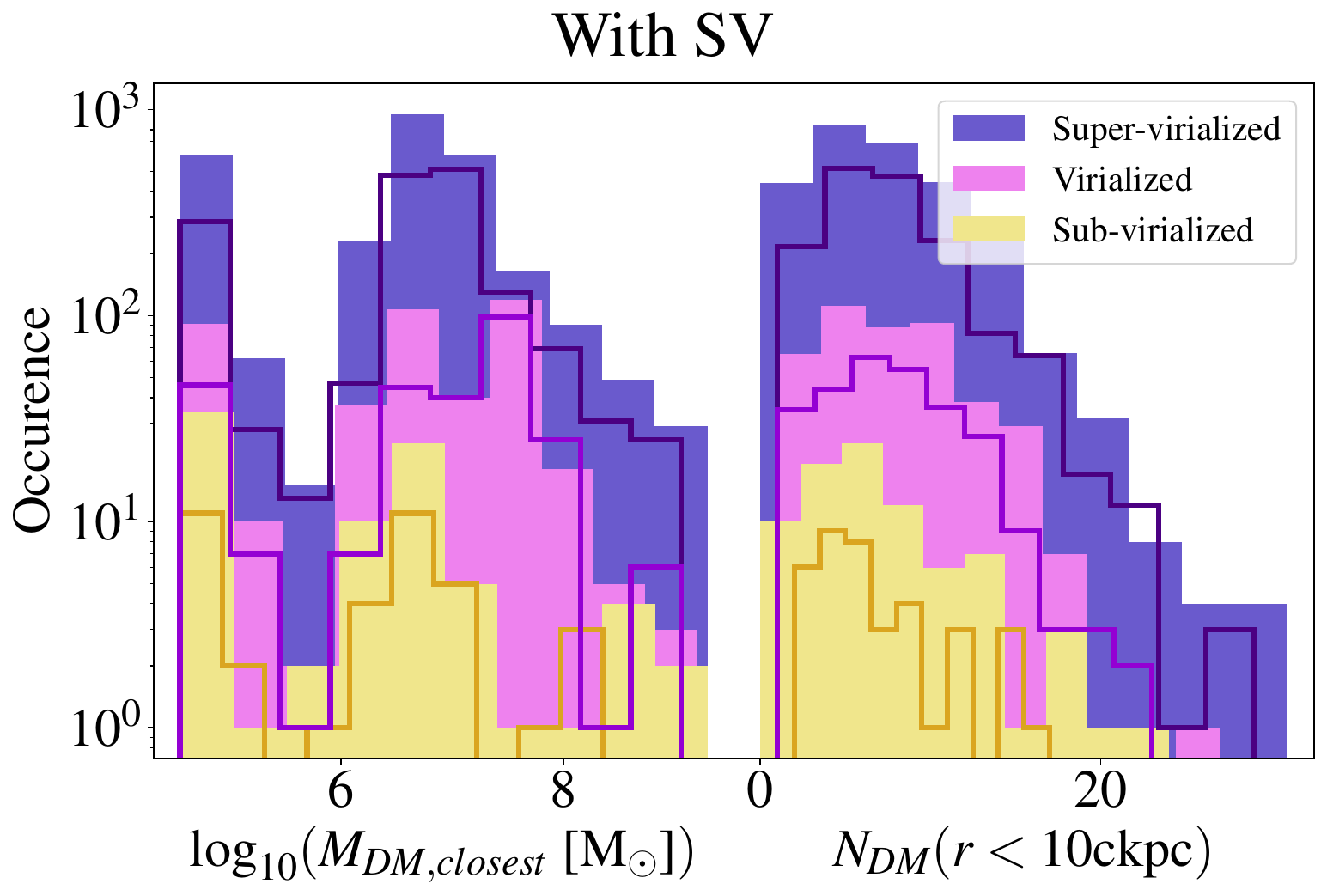}

    \caption{{\bf Dark matter halo nearby neighbors:} Left panels: Histogram of the logarithm of the dark matter mass of the nearest dark matter halo to star clusters in the stars+gas centric catalog. Filled in histograms show $z=12$, and the overplotted lines show $z=15$. Right panels: Number of dark matter halos within 10 ckpc of the center of mass of star clusters. Panels depict super-virialized systems (dark blue), virialized systems (pink) and sub-virialized systems (yellow). The top row shows results without the stream velocity, and the bottom row shows results with the stream velocity.}
    \label{fig:dmobj_properties}
\end{figure}

\begin{figure}
    \centering
    \includegraphics[width=0.7\linewidth]{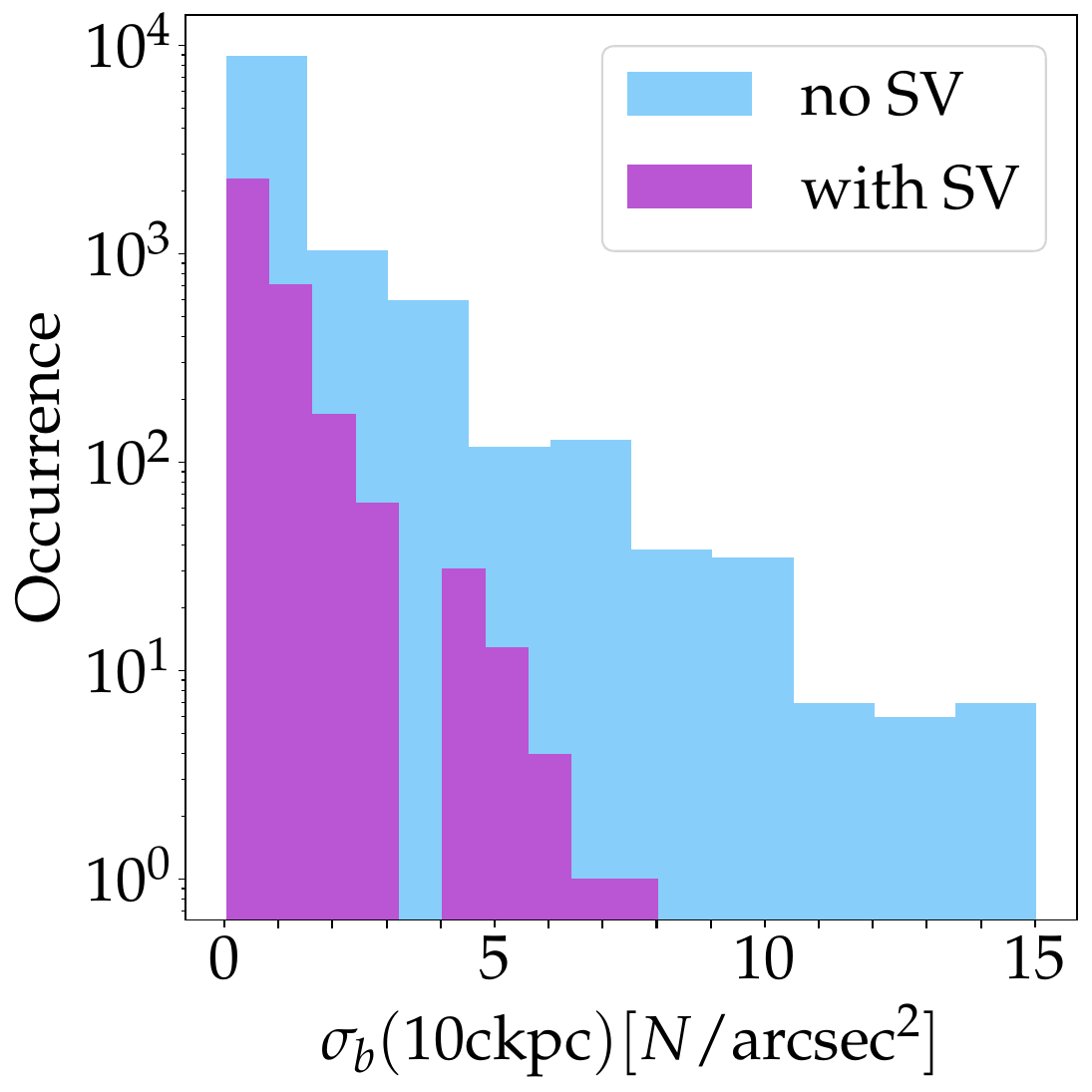}
    \caption{{\bf Baryonic structure neighbors}
    Occurrence of baryonic objects by the surface density of baryonic structures in the surrounding 10~ckpc region (per square arcsec). 
    }

    \label{fig:baryobobj_neighbors}
\end{figure}

One of the benefits of the method used here to distinguish both dark matter halos and star clusters using two FOF algorithms is that it allows us to probe the environment and structure of stellar systems.
We explore the nature of the population of nearby neighbors to our star clusters in Figs.~\ref{fig:dmobj_properties}-\ref{fig:baryobobj_neighbors}. 
In Fig.~\ref{fig:dmobj_properties}, we investigate the nearest neighbor to each baryonic star cluster object in the dark-matter centric catalog. 
The plot shows histograms of the mass of the closest halo to the star cluster (which may be its host) and the number of neighbors within 10 ckpc. 
The population of objects is split into super-virialized, virialized, and sub-virialized systems.
The plot shows objects at $z=15$ (solid lines) and $z=12$ (filled-in histograms).
Compared to objects forming in the simulation box without the stream velocity, star clusters with the stream velocity (bottom plot) are likelier to have a more massive nearest/host halo. 
This reflects the reduced gas fraction in stream velocity halos, causing a less massive star cluster to form for the same halo mass. 
Additionally, at both redshifts, very few star clusters are found in the stream velocity simulation that are both virialized and have many (15-20+) neighbors. 
This seems to be a result of the delayed onset of structure formation due to the stream velocity effect. 
Systems form at a later redshift, and thus have had less time to tend towards dynamical equilibrium, especially in complex, clustered systems where many low mass halos interact. 

As discussed in the previous section, without the stream velocity, star clusters are more likely to be in a region containing many other star clusters, which contributes to the total luminosity of these stellar complexes.
In Fig.~\ref{fig:baryobobj_neighbors}, we plot the occurrence of systems by the baryon object surface density in their nearby environment. 
This is found by averaging the number of neighbors within 10 ckpc (similar to the right panel of Fig.~\ref{fig:dmobj_properties}). 
We report the density in terms of square arcsec to give a sense of the occurrence on the sky. 
With the stream velocity, the number of systems in regions of five objects per square arcsec is strongly reduced.
Note that this is already in an extremely high density region, as generated by our high $\sigma_8$. 
Thus, if a Pop III star cluster is discovered through lensing or another scenario and is shown to have many sub-clumps, we suggest that it is highly unlikely to be located in region where the stream velocity is $\sim 2\sigma_{\rm bc}$. 

\subsection{Dynamical state of star clusters}
\label{subsec:dynamicalstate}

\begin{figure}
     \centering
    \includegraphics[width=0.95\linewidth]{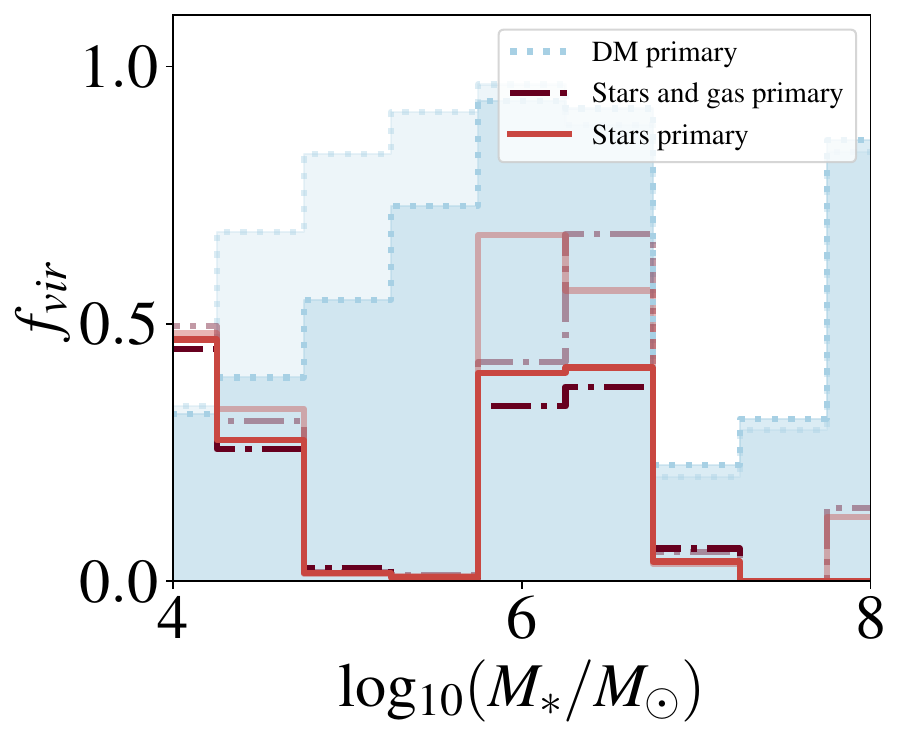}
    \caption{Fraction of virialized objects versus stellar mass in the box  with stream velocity (solid) and without (faint/transparent histograms). The line styles follow the convention of Fig.~\ref{fig:mstar_hist}.}
    \label{fig:virialized_hists} 
\end{figure}

\begin{figure}
    \centering
    \includegraphics[width=0.9\linewidth]{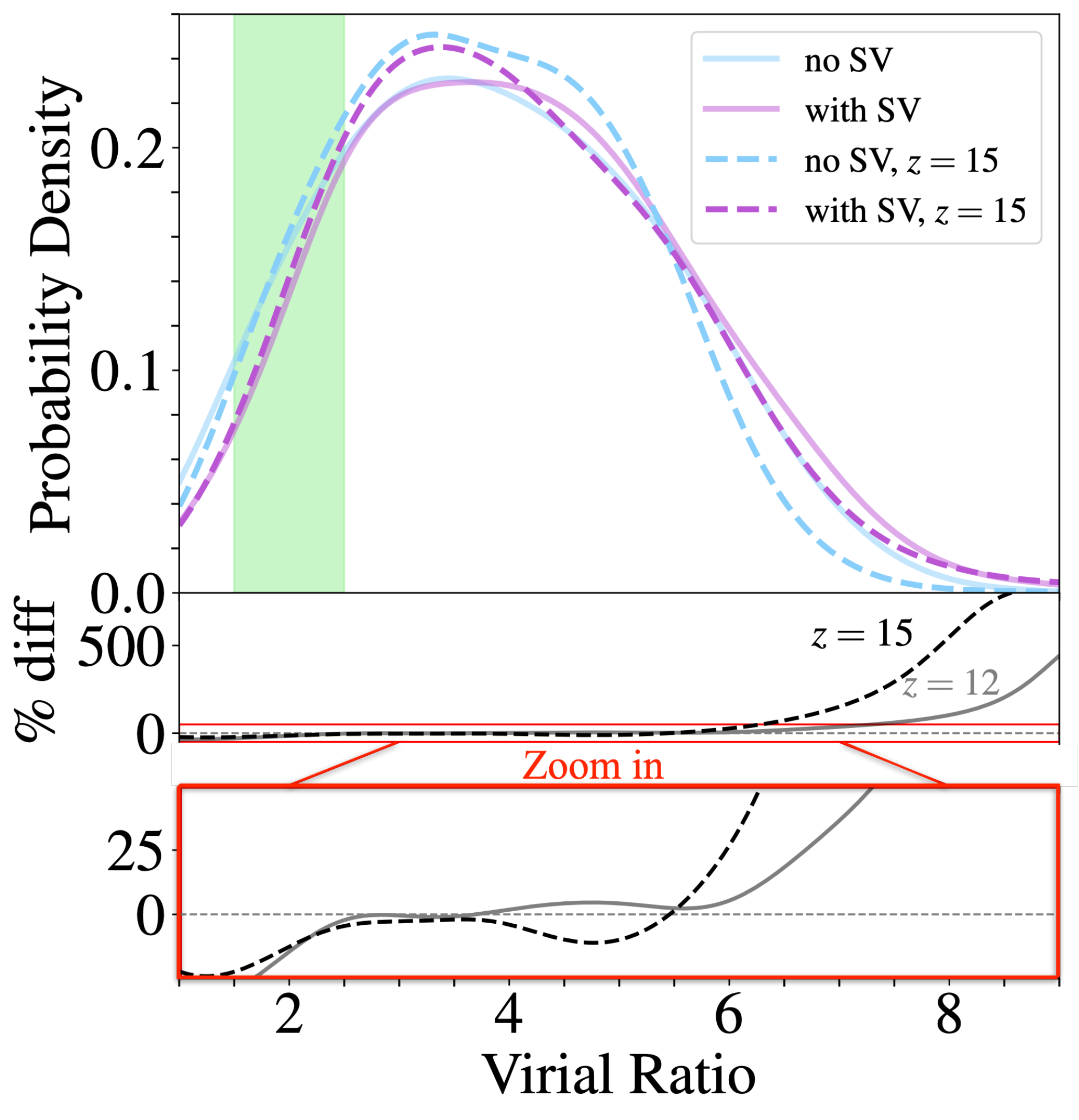}
    \caption{Probability density distribution of virial ratios at $z=12$ (solid lines) and $z=15$ (dashed). Objects in the stream velocity simulation are shown in light blue, and objects in the box without the stream velocity are shown in purple.
    The bottom panel shows the percent difference between the stream velocity and no stream velocity curves at each redshift. 
    The green region shows the systems that meet our threshold for virialization. Probability densities are estimated using a Gaussian kernel density with bandwidth of 0.35. }
    \label{fig:virial-ratios}
\end{figure}

The virialization state of star clusters at high redshift is a balance between the trend towards equilibrium over time and the disruption from interaction with nearby halos. 
Using the simple dynamical arguments presented in \cite{Williams+25}, whether or not a system is virialized can be described by comparing two timescales, the virialization time (${t_{\rm vir}}$) and the interaction time ($t_{\rm int}$), to the age of the Universe. 
Once the virialization timescale (which is dominated by the cooling time at low masses) drops below the age of the Universe, systems should achieve virialization as they tend towards dynamical equilibrium. 
In the simulation box, the stellar system is likely to be found in this virialized state as long as the interaction timescale for major mergers is longer than the virialization timescale.
This is true for high density systems if $M_*\lesssim10^{5}M_\odot$ and low density systems with $10^6M_\odot\lesssim M_*\lesssim 10^{7} M_\odot $.
Above $10^7M_\odot$,  interaction with massive halos becomes common. 
Finally at $10^8 M_\odot$,  
systems have generally achieved virialization since few similarly massive perturbing halos exist in the box. 
Here, we investigate whether this framework applies to the regions with the stream velocity or not, given that supersonic streaming affects the density of halos, the turbulence of gas, and the morphology of structures \citep[e.g.,][]{Williams+23,Hirano+23,chen_supersonic_2025}.

In Figure \ref{fig:virialized_hists}, we compare the counts of virialized objects by stellar mass in the box with the stream velocity, plotting the fraction that are virialized at each mass.
The no-streaming distribution is shown as transparent histograms.
From the red histograms denoting the baryonic catalogs, we can see that both boxes agree with a split into two populations of virialized objects, located at roughly the same masses.
A dearth of virialized systems is present with and without the stream velocity at $\sim10^5M_\odot$
This indicates that the description provided in \cite{Williams+25}, which depends on the cooling and dynamical timescales, rather than the stream velocity, captures the physical behavior of the star cluster dynamics well enough to describe the location of the virialization peaks. 

The stream velocity plays a role in the baryonic catalogs' distributions around $M_*\sim 10^{6-7}M_\odot$ in particular. 
With the stream velocity, a smaller fraction of systems ($\sim40\%$) in the high-mass subset are virialized than without the stream velocity, where the fraction is around $70\%$. 
This may indicate that the higher-mass systems in regions of streaming may take slightly longer to achieve virialization. 
This delay likely stems from the same effect that delays the onset of Pop III star formation with the stream velocity, as seen in many studies  \citep[e.g.,][]{Schauer+19, Lake+24a, Williams+24}. 
For these systems, initial gas accretion into primordial halos is delayed due to the stream velocity's advection of the gas from the halo, leading to delays in their further evolution.


To investigate this, we explore the evolution of clusters' virial ratios with redshift. 
Figure~\ref{fig:virial-ratios} shows probability density distributions of the virial ratios at $z=15$ and $z=12$. 
The green highlighted region shows our virialization criterion,

\begin{equation}
    1.5\le -U/K\le 2.5 \ .
\end{equation}

The distribution peaks at a slightly super-virial value between $3-4$, with a significant fraction of systems having extremely high virial ratios. The super-virial relation persists until the end of the simulation at $z=12$.  This excess of gravitational potential energy may suggest that the continued evolution of these systems will be towards an even more compact configuration. In other words, in the absence of dynamical processes that may stabilize these clusters, they may continue to collapse. 

The $z=15$ stream velocity run displays an excess of objects with extremely high virial ratios compared to the no stream velocity case. 
However, by $z=12,$ the stream velocity systems approach the non-streaming distribution. 
These may reflect systems that formed later due to the stream velocity and thus are strongly out of equilibrium in the earlier snapshot. 
At this later time, these systems have had time to trend towards the typical dynamical path of violent relaxation and environmental disruption seen in the non-streaming box.

\section{Discussion}
\label{sec:summary}
In this work, we provide a more comprehensive picture of dense star cluster formation within the $\Lambda$CDM framework. 
We show that, given a no feedback scenario, even regions with high stream velocity values still produce star clusters with sufficiently high stellar surface densities to be consistent with the JWST observed population. 
We investigate these high-stream-velocity regions  using a methodology focused on identifying baryonic structures, which enables us to examine their star cluster dynamics and environment in detail. 

\paragraph{Effects of baryonic feedback}
In our simulations without baryonic feedback, the number of brighter objects in our no-stream-velocity simulations is boosted compared to the estimate in \cite{Williams+24} due to the inclusion of star clusters that were undetected using dark matter focused structure finding algorithms. This includes many objects which live in halos that are not resolved or detected by these algorithms. Our simulations suggest that these no-streaming systems are clumpy, and are potentially composed of a number of merged star clusters that were once their own dwarf galaxies, or may also be forming many smaller clumps in place. We suggest that the inclusion of feedback may disrupt this small-scale mode of star formation. As a result, the original discrepancy noted in \cite{Williams+24} might reappear even when using the baryonic-focused identification method with feedback incorporated.

One important consideration regarding feedback effects is how these processes may vary between regions with stream velocity and those without it. 
Although the literature does not suggest that stream velocity directly influences the physics of stellar feedback, differences in geometry and environmental conditions might lead to different scenarios.
For instance, the clustering of stellar systems can play a crucial role. As shown in Fig.~\ref{fig:baryobobj_neighbors}, non-stream velocity systems are significantly more likely to develop in areas with a high concentration of nearby stellar systems. 
In these dense regions, star formation may initiate sporadically; for example, when star formation begins in one clump, it can inhibit the ability of nearby gas to form additional star clusters. 
This aspect is not accounted for in our no-feedback simulation, meaning that star formation could be considerably suppressed or even entirely prevented, resulting in a reduced number of clusters in regions without stream velocity. 
Conversely, systems characterized by stream velocity, which tend to be more isolated, would be less impacted by this effect.

\paragraph{Metal enrichment}
Pristine, primordial gas, found in early minihalos, can become metal-enriched on the timescale of the lifetime of massive stars. 
In our simulations, no metal enrichment is included. 
As discussed in \S~\ref{subsec:detectability}, many of the more massive objects in the simulation box are found in an environment with other stellar clumps, many of which are no longer forming stars at an elevated rate. 
This suggests a previous burst of star formation, which can enrich some portion of the gas, causing observable metal lines. 
A striking difference in the stream velocity systems is the reduced number of neighbors in the local environment, as discussed in \S~\ref{subsec:environment}. 
Many studies have investigated the delay in the initial onset of star formation caused by the stream velocity, noting that it may prolong the epoch of first star formation, affecting the observability of Pop III stars  \citep[e.g.,][]{Stacy+10,Greif+11,Schauer+17a,Schauer+21, Schauer+23,Naoz+12, BD, Asaba+16,Nebrin+23,Hegde+23,Conaboy+23}. 
Here, we suggest that the difference in environmental clustering is an additional effect that may serve to boost the continued presence of Pop III star clusters at lower redshift. 
With the stream velocity systems form later, leading to Pop III stellar populations at slightly later times. 
However, we show here that once these Pop III star clusters form, they typically form alone in a burst, with fewer nearby systems that may have pre-enriched the natal gas. 
Thus, we speculate that gas metallicity in and around the earliest low mass structures traces stream velocity induced enrichment. 
This may appear through the stellar metallicity and nebular metallicity, thanks to a delayed progression of local enrichment, as well as the metallicity of eventual outflowing gas, kicked out by Pop III feedback processes. 
Further studies that resolve the disbursement of metals into the surrounding medium are needed to explore this question \citep[e.g.,][]{mead_aeos_2025}.

\begin{figure}
    \centering
    \includegraphics[width=0.95\linewidth]{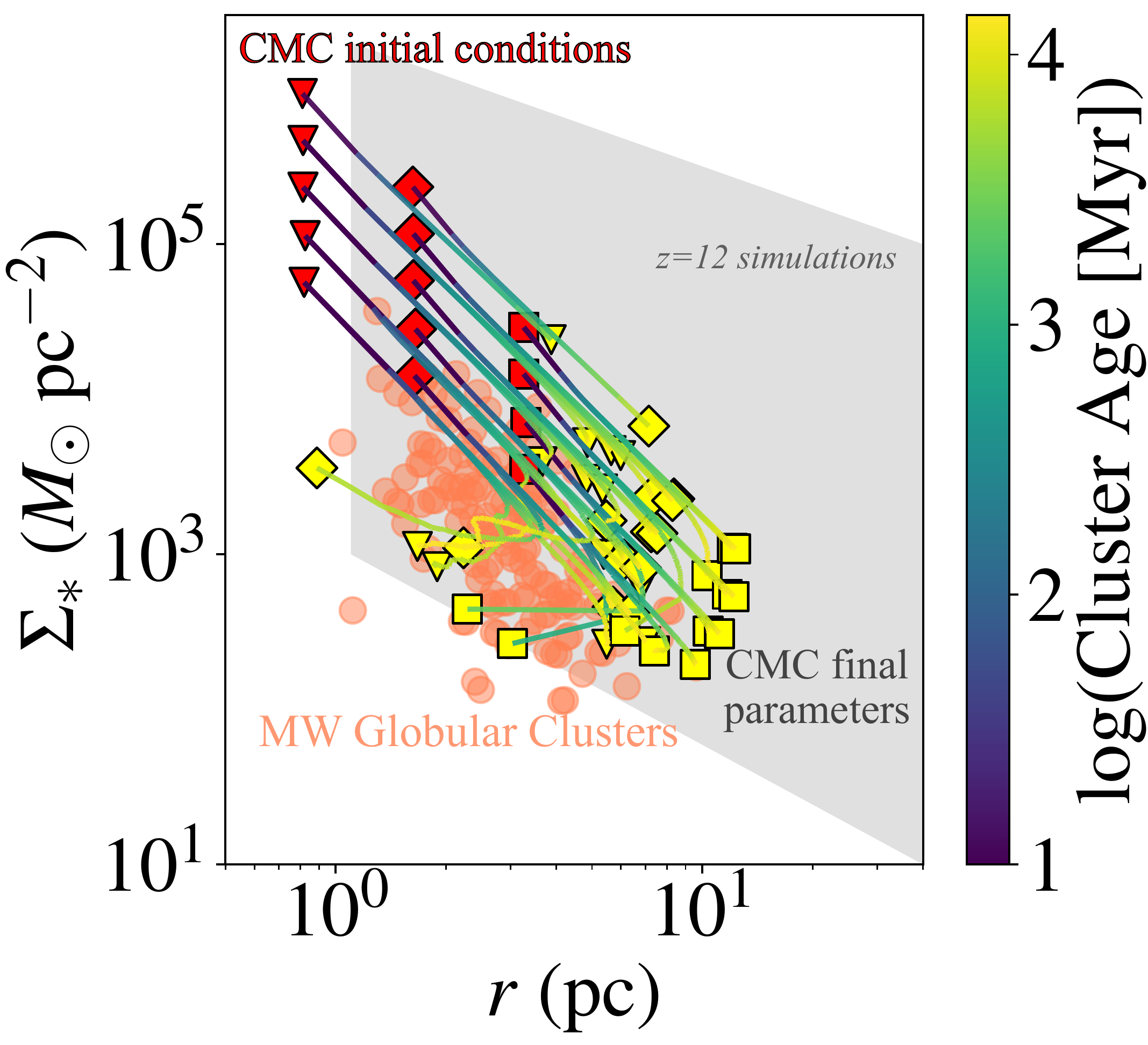}
    \caption{Comparison of $z=12$ star clusters from this work (grey shaded region encompassing the points from Fig.~\ref{fig:surface_density}) with evolutionary tracks of N-body simulations of globular cluster evolution with the Cluster Monte Carlo (CMC) code \citep[][color bar and red/yellow scatter]{kremer_modeling_2020}, and Milky Way globular clusters with $\Sigma_*>10^4 M_\odot \text{pc}^{-2}$ \citep[][coral circles]{Gieles+11}. The plot shows surface density versus radius, as in Fig.~\ref{fig:surface_density}. 
    The color bar shows the age of the cluster evolutionary track.
    Red scatter points show the CMC initial conditions, and yellow scatter show the clusters' final parameters at the end of the CMC evolution. The scatter points are shaped according to initial radius.
    We use the subset of evolutionary models from \cite{kremer_modeling_2020} with $Z=0.0002$ and $r_{\rm h, init}>0.7$pc.
    The tracks differ based on initial number of particles, virial radius, and galactocentric distance. }\label{fig:CMCsurface_density}
\end{figure}

\paragraph{Evolution of star clusters to low redshift}
While our simulation's final snapshot is taken at $z=12$, it is interesting to consider the continued evolution of the stellar objects in our simulation box. 
To estimate what the properties of these dense star clusters may be today, we compare our objects to the N-body simulations of \cite{kremer_modeling_2020}. 
Shown in Fig.~\ref{fig:CMCsurface_density}, these evolutionary tracks cover $10-13$ Gyr of clusters' dynamical evolution, for a variety of masses, virial radii, and galactocentric distances. 
These simulations follow the evolution of individual clusters, without a cosmological box, so interaction with a broader environment is not included other than effective tidal mass loss from galactic tides at a set distance. 
For the most representative comparison to our simulation, we select the lowest metallicity subset of models ($Z=0.0002$), and those with initial half-mass radius $r_{\rm h, init}>0.7$ pc.
These CMC model initial parameters do not cover the full range of objects in our simulation--which represent a range of clusters and protogalaxies. In particular, the low surface density and high mass objects do not have analogs in the CMC catalog. However, clusters with comparable surface density and size to some of our high density, lower mass systems are interesting to consider. 

The Figure demonstrates that clusters with an initially very high surface density between $10^3-10^5 M_\odot$ pc$^{-2}$ evolve over time to higher radii and lower surface density. 
Some of the clusters eventually experience core collapse. 
Comparing our objects to the evolutionary tracks of CMC clusters, we see that a subset of our smallest simulations occupy a region of parameter space where the precursors of observed Milky Way globular clusters lie. 
We suggest that following the end of star formation, the high density subset of objects we simulate at Cosmic Dawn may evolve to become low metallicity, high density globular clusters at the present time, and may accrete onto dwarf and ultra-faint galaxies.  

\section{Summary}
In summary, we show that Pop III star clusters forming in a $\Lambda$CDM cosmological box can achieve extremely high surface density consistent with {\it JWST} observations, even when the stream velocity is included. 
Our careful accounting for systems' observable properties--through a baryonic-focused structure-finding algorithm and association of unresolved complexes, are used in combination with a semi-analytic approach to estimate observable properties.  
We estimate that the brightest of the systems in our box are detectable as the faintest current {\it JWST} galaxies, but that lensing scenarios could reveal low mass objects, especially those whose luminosity is heightened due to IMF effects or bursty star formation in its initial phases. 
Additionally, the boost in the faint-end UVLF in the stream velocity versus the non-stream velocity case seen when counting by dark matter halos is obfuscated in the context of the faint-end UVLF. 
However, the suppression at extremely faint magnitudes is still preserved, and additionally the statistical clumping of small structures may be a useful probe into the effects of the stream velocity at high redshift.

\section*{Acknowledgements}

 The authors would like to thank Leonardo Clarke for insightful conversations. 
 C.E.W.  acknowledges the support of the National Science Foundation Graduate Research Fellowship, the University of California, Los Angeles (UCLA), the Mani L. Bhaumik Institute for Theoretical Physics, and the UCLA Center for Developing Leadership in Science Fellowship. C.E.W., W.L., S.N., Y.S.C, B.B., F.M., and M.V. thank the support of NASA grant No. 80NSSC24K0773 (ATP-23-
ATP23-0149) and the XSEDE/ACCESS AST180056 allocation, as well as the UCLA cluster Hoffman2 for computational resources. C.E.W and S.N. thank Howard and Astrid Preston for their generous support. B.B. also thanks the the Alfred P. Sloan Foundation and the Packard Foundation for support.
N.Y. and C.E.W. acknowledge financial support from JSPS International Leading Research 23K20035.
 F.M. acknowledges support by the
European Union—NextGeneration EU within PRIN 2022
project n.20229YBSAN—Globular clusters in cosmological
simulations and in lensed fields: from their birth to the present
epoch.
 Simulation runs for this work were performed on the Anvil Cluster \citep{song_anvil_2022}, and C.E.W. thanks ACCESS Discover project PHY250057 for computational resources. 
This material is based upon work supported by the National Science Foundation Graduate Research Fellowship Program under Grant No. DGE-2034835. Any opinions, findings, conclusions, or recommendations expressed in this material are those of the author(s) and do not necessarily reflect the views of the National Science Foundation.
This work used computational and storage services associated with the Hoffman2 Cluster which is operated by the UCLA Office of Advanced Research Computing’s Research Technology Group.

\software{astropy \citep{2013A&A...558A..33A,2018AJ....156..123A}, matplotlib \citep{Matplotlib},  numpy \citep{harris2020numpy}, scipy \citep{2020SciPy-NMeth}.}

\bibliography{cosmo}{}
\bibliographystyle{aasjournal}

\appendix 
\section{Catalog construction}
\label{App:catalog}
\begin{figure}
    \centering
    \includegraphics[width=0.5\linewidth]{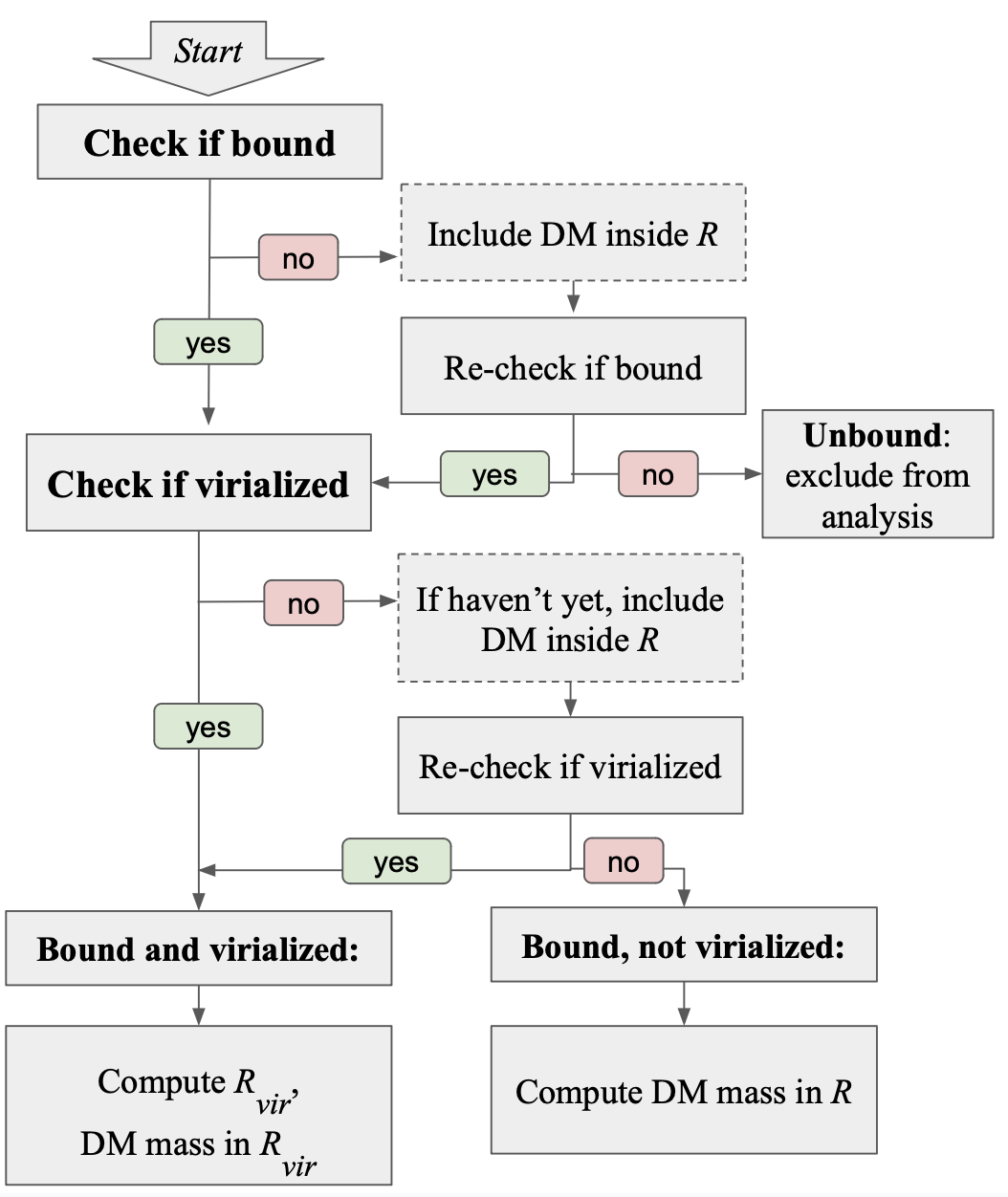}
    \caption{Flowchart describing the calculation of the boundedness and virialization of objects. Reproduced from Fig.~1 of \cite{Williams+25}. }
    \label{fig:flowchart}
\end{figure}

Figure~\ref{fig:flowchart} depicts the process used to eliminate unbound structures from our catalog. 
First, the kinetic and potential energy of all star particles associated with each object is calculated. 
If the stars are not bound, the procedure is repeated, including the additional contribution of the dark matter particles within the radius of the object. 
If the objects fails to be gravitationally bound after this second stage of calculation, we exclude if from further analysis. 
A similar procedure is used to test for the systems' dynamical state, but both virialized and non-virialized systems are included in the final catalog.

\section{Convergence}
\label{app.conv}
Here we test our cutoff of 100 star particles to filter out objects from the baryonic catalogs which we do not numerically resolve. 
In particular, we seek to ensure that our categorization of the virial ratio is robust, even at low particle number. 
This question was already investigated in \cite{Williams+25} for the no stream velocity box, but here we replicate the analysis for the box including the stream velocity. 

\begin{figure}
    \centering
    \includegraphics[width=0.5\linewidth]{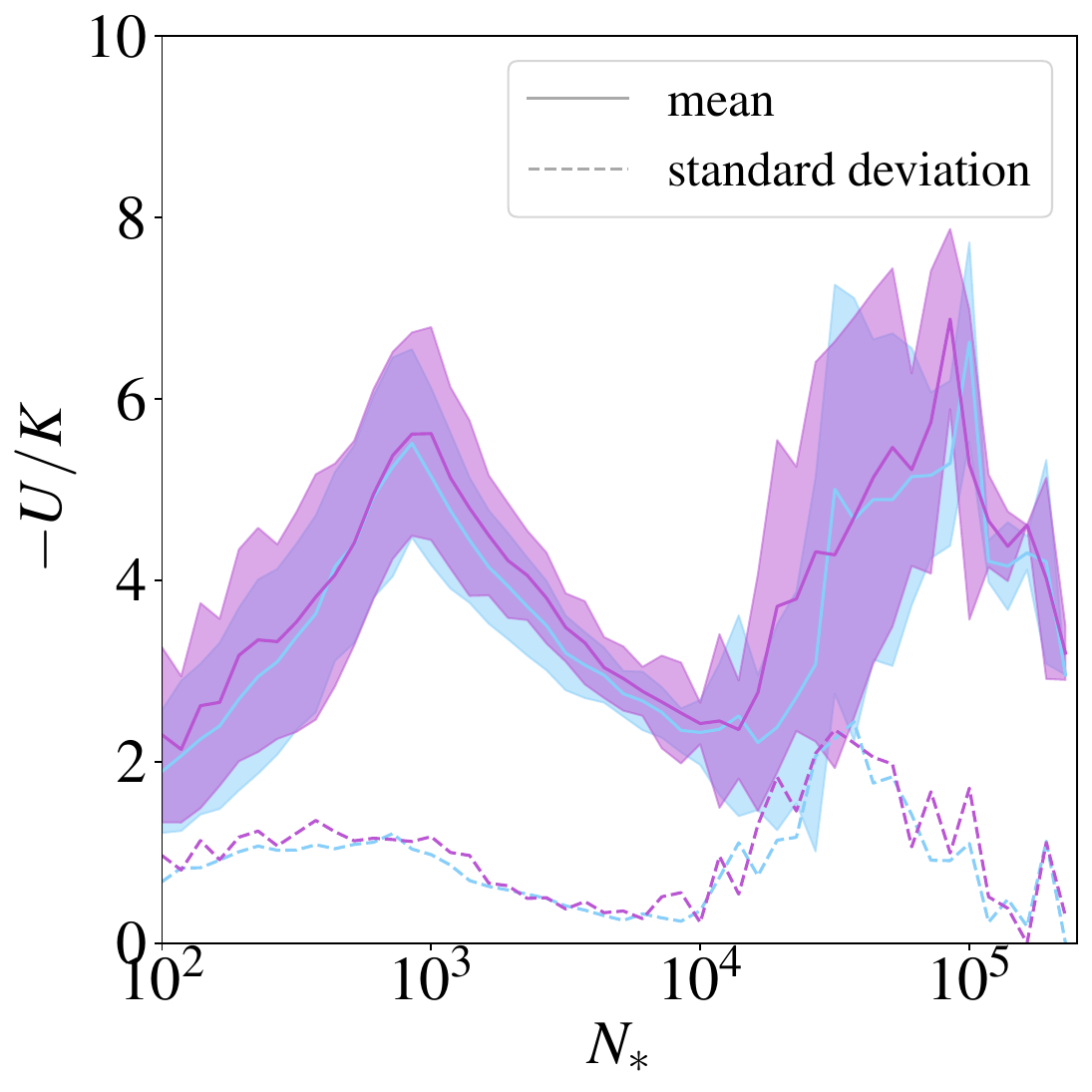}
    \caption{Virial ratio versus number of star particles for objects in the stars and gas primary catalogs with $v_{\rm bc}= 0\sigma_{\rm bc}$ (blue) and $v_{\rm bc}= 2\sigma_{\rm bc}$ (purple). The means are shown as solid lines and the standard deviation is shown as the dashed line. The shaded regions show one standard deviation above and below the mean. }
    \label{fig:virratio_nstars}
\end{figure}

In Fig.~\ref{fig:virratio_nstars}, we plot the virial ratio as a function of $N_*$, the number of star particles. 
The mean is shown, with shaded regions indicating one standard deviation above and below the curve. 
Additionally, the value of the standard deviation is shown with the dashed line. 
From the plot, we can see that for both catalogs, the variance in the virial ratio is generally low ($\sim 1$), and remains roughly constant over an order of magnitude in particle number. 
Thus, the dynamical state of objects is robustly identified. 

\section{Supplemental data}
\label{App:suppplemental}
\begin{figure*}
    \centering
    \includegraphics[width = 
    0.8\textwidth]{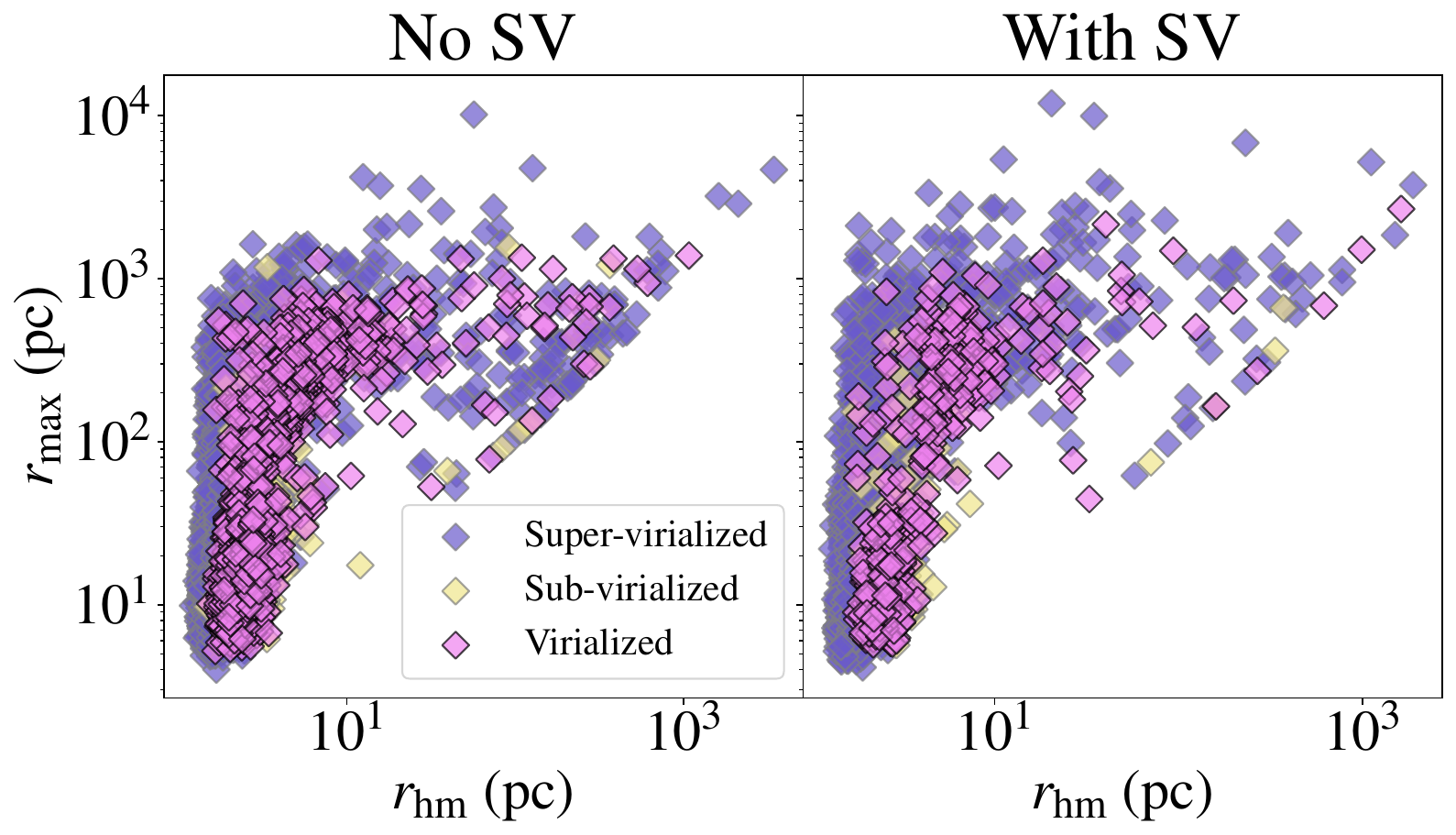}
    \includegraphics[width = 
    0.8\textwidth]{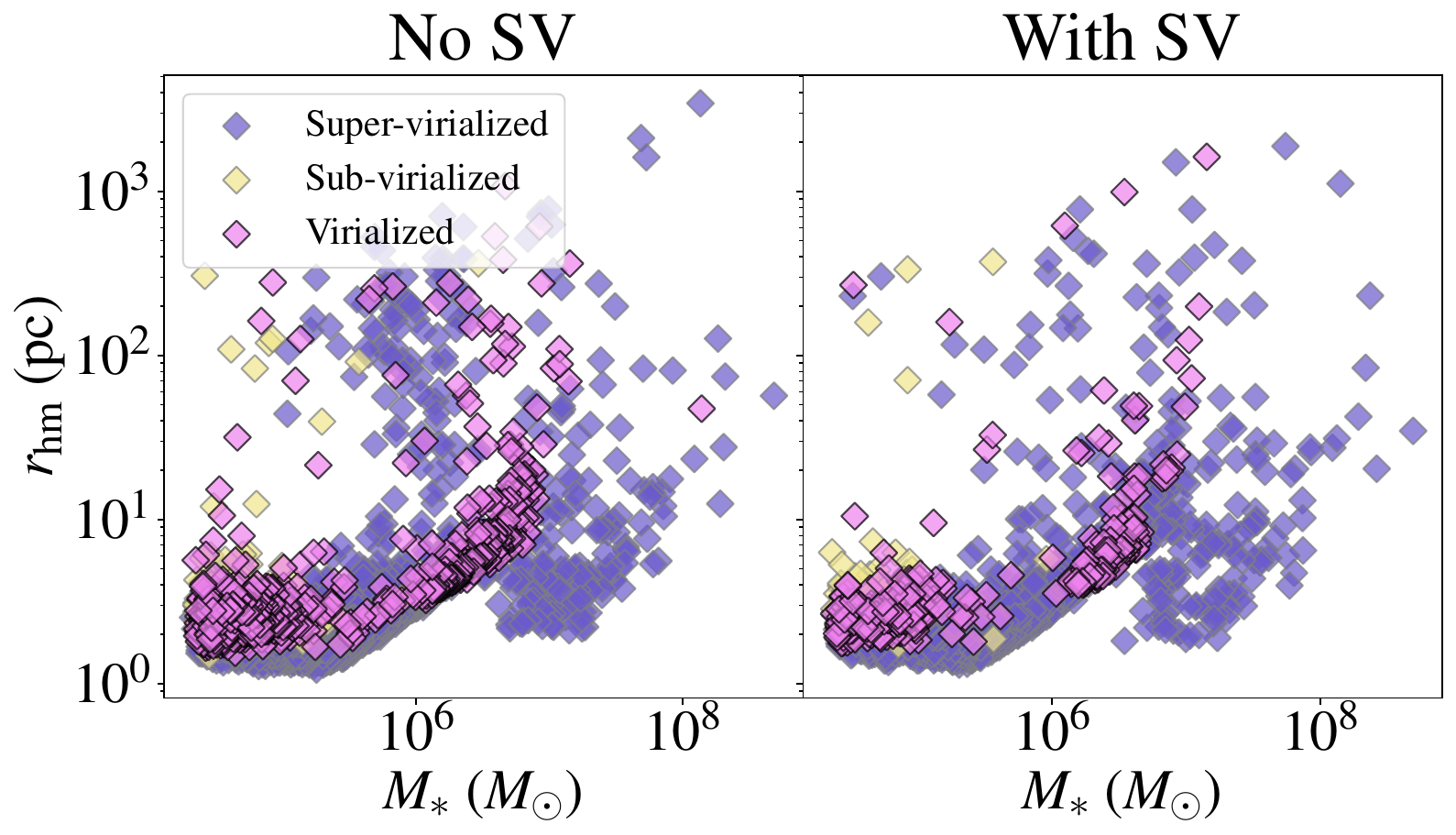}
    \caption{Top panels: Maximum radius versus half mass radius of star clusters. 
    Bottom panels: Half mass radius versus stellar mass of star clusters. 
    The left hand side of each panel shows the run without stream velocity, while the right hand side shows the run with the inclusion of the stream velocity. The dark purple scatter shows the super-virialized points, the yellow shows sub-virialized objects, and the pink shows the virialized objects. Results are presented at $z=12$ for the stars $+$ gas primary run. }
    \label{fig:radii}
\end{figure*}

In Fig.~\ref{fig:radii}, we provide scatter plots of the radii associated with objects in the baryonic-focused catalog, with and without the stream velocity. 
The top panel shows the maximum radius versus half-mass radius relation, and the bottom panel shows the half mass radius as a function of stellar mass. 

\end{document}